\documentclass[twocolumn,showpacs,showkeys,nofootinbib,superscriptaddress]{revtex4}
\usepackage{graphicx}
\usepackage{amsmath,amsfonts,amssymb,bm}

\begin{document}

\title{Muon capture by ${}^{3}\mathrm{He}$ nuclei followed by proton
and deuteron production}

\author{V.M.~Bystritsky}
\altaffiliation{Corresponding author}
\email{bystvm@nusun.jinr.ru}
\affiliation{Joint Institute for Nuclear Research, Dubna 141980,
Russia}

\author{V.F.~Boreiko}
\affiliation{Joint Institute for Nuclear Research, Dubna 141980,
Russia}

\author{M.~Filipowicz}
\affiliation{University of Mining and Metallurgy, Fac.~of Fuels and
Energy, PL--30059 Cracow, Poland}

\author{V.V.~Gerasimov}
\affiliation{Joint Institute for Nuclear Research, Dubna 141980,
Russia}

\author{O.~Huot}
\affiliation{Department of Physics, University of Fribourg, CH--1700
Fribourg, Switzerland}

\author{P.E.~Knowles} 
\affiliation{Department of Physics, University of Fribourg, CH--1700
Fribourg, Switzerland}

\author{F.~Mulhauser}
\affiliation{Department of Physics, University of Fribourg, CH--1700
Fribourg, Switzerland}

\author{V.N.~Pavlov}
\affiliation{Joint Institute for Nuclear Research, Dubna 141980,
Russia}

\author{L.A.~Schaller}
\affiliation{Department of Physics, University of Fribourg, CH--1700
Fribourg, Switzerland}

\author{H.~Schneuwly}
\affiliation{Department of Physics, University of Fribourg, CH--1700
Fribourg, Switzerland}

\author{V.G.~Sandukovsky}
\affiliation{Joint Institute for Nuclear Research, Dubna 141980,
Russia}

\author{V.A.~Stolupin}
\affiliation{Joint Institute for Nuclear Research, Dubna 141980,
Russia}

\author{V.P.~Volnykh} 
\affiliation{Joint Institute for Nuclear Research, Dubna 141980,
Russia}

\author{J.~Wo\'zniak}
\affiliation{University of Mining and Metallurgy,
Fac.~Phys.~Nucl.~Techniques, PL--30059 Cracow, Poland}

\date{\today}

\begin{abstract}
The paper describes an experiment aimed at studying muon capture by
${}^{3}\mathrm{He}$ nuclei in pure ${}^{3}\mathrm{He}$ and
$\mathrm{D}_2 + {}^{3}\mathrm{He}$ mixtures at various densities.
Energy distributions of protons and deuterons produced via
$\mu^-+{}^{3}\mathrm{He}\to p+n+n + \nu_{\mu }$ and $\mu^-+{}^{3}
\mathrm{He} \to d+n + \nu_{\mu }$ are measured for the energy
intervals $10 - 49$~MeV and $13 - 31$~MeV, respectively.
Muon capture rates, $\lambda _\mathrm{cap}^p (\Delta E_p )$ and
$\lambda _\mathrm{cap}^d (\Delta E_d )$ are obtained using two
different analysis methods.
The least--squares methods gives $\lambda _\mathrm{cap}^p = (36.7\pm
1.2) \, \mbox{s}^{ - 1}$, $\lambda_\mathrm{cap}^d = (21.3 \pm 1.6)
\,\mbox{s}^{ - 1}$.
The Bayes theorem gives $\lambda _\mathrm{cap}^p = (36.8 \pm 0.8) \,
\mbox{s}^{ - 1}$, $\lambda _\mathrm{cap}^d = (21.9 \pm 0.6)
\,\mbox{s}^{ - 1}$.
The experimental differential capture rates, $d\lambda _\mathrm{cap}^p
(E_p ) / dE_p $ and $ d\lambda _\mathrm{cap}^d (E_d ) / dE_d$, are
compared with theoretical calculations performed using the plane--wave
impulse approximation (PWIA) with the realistic NN interaction Bonn B
potential.
Extrapolation to the full energy range yields total proton and
deuteron capture rates in good agreement with former results.
\end{abstract}


\maketitle

\section{Introduction}
\label{sec:introduction}

The study of few--nucleon systems is interesting and very
important. 
It gives a microscopic description of complex systems within the
framework of modern concepts of nucleon--nucleon
interaction~\cite{kharc92}.
Using the nuclear muon capture to study few--nucleon systems is
a perfect tool since the nuclear structure had been found to play an
important role in such systems~\cite{balas98,measd01}.
Energy transferred to a nucleus when muon capture occurred causes the
excitation of low--lying levels in the residual nucleus up to the
giant resonance region~\cite{foldy64} or emission of
intermediate--energy neutrons~\cite{mukho77}.
This picture is clear within the framework of the plane--wave impulse
approximation (PWIA)~\cite{dautr76} (and references therein).
However, some experiments~\cite{measd01,kozlo85,schaa83,macin84}
indicate that the energy transferred to the residual nucleus in muon
capture is large.
It was found in those experiments that collective nuclear excitations
like giant resonances play a decisive role in the muon capture
process.
In most cases the decay of the giant resonance was followed by the
emission of a neutron and the formation of a daughter nucleus in the
above--threshold state for which it was then ``beneficial'' to decay
via the proton or deuteron
channel~\cite{kozlo85,schaa83,macin84,pluym86,leexx87}.

An interesting feature of such nuclear decays is the emission of
high--energy ($40 - 70$~MeV) charged particles (protons,
deuterons)~\cite{budya71,baldi78,krane79,belov86,kozlo78,marto86}.
By studying such an emission resulting from nuclear muon capture it is
possible to get information both on the nuclear structure and the muon
capture mechanism itself~\cite{balas98,measd01}.
The emission of high--energy protons and deuterons in muon capture
seems to be due to the existence of initial or final--state nucleon
pair correlations and to a contribution to the interaction from the
meson exchange currents (MEC)~\cite{lifsh88,berna77}.
Note that the MEC contribution is very sensitive to the details of the
wave function for the nuclear system.

In the region of large energy transfer (extreme kinematics case) the
MEC contribution to the interaction becomes substantial.
Note that MEC and nucleon--nucleon correlation effects are included
``automatically''.
For example, the calculation of the rate for muon capture by a
deuteron~\cite{goula82,doixx90} indicates that inclusion of MEC in the
muon capture matrix element considerably increases the calculated
capture rate at the boundary of the kinematic region as compared to
the contribution from the high--momentum components of the deuteron
wave function.
The above mentioned factors may cause nuclear transitions with a large
energy transfer.

Though yields of charged particles in the muon capture process are
relatively small, the study of these events may give more information
than other methods: it provides an insight into the mechanism for
excitation and decay of nuclei upon muon capture.
So far, there is no microscopic description of the nuclear muon
capture process~\cite{balas98}.
To ensure a correct comparison between theory and experiment, it is
necessary to study muon capture in few--nucleon systems ($A\le 3$),
where a microscopic calculation of wave functions in the initial and
final states is possible~\cite{goula82,doixx90}.

Matrix elements calculations for the nuclear muon capture transitions
are usually performed using the wave functions model of the initial and
final states.
The wave function parameter values are chosen such that calculated and
experimental data agree correctly for the case of low--lying nuclear
states spectra and corresponding magnetic moments~\cite{balas98}.
In the case of light nuclei a multi particle shell model is frequently
used.
This model describes (with a defined accuracy) these
characteristics, i.e., the spectra and magnetic moments.
However, the shell model accuracy may become insufficient because of
poor knowledge of muon--nucleon interaction constants.
In addition, there remains the problem of MEC.

At present, general properties of nuclear transitions to the
continuous spectrum for muon capture are treated on the basis of a
resonant collective mechanism for the muon absorption by a
nucleus~\cite{balas98,measd01}.
The strongest $E1$ transitions, much like nuclear photo-disintegration
reactions, form a giant dipole resonance and are collectivized into a
continuous spectrum at muon capture~\cite{balas98}.
The character of collective motions excited in nuclei at muon
absorption is different from that in nuclear photo-disintegration
reactions.

The giant resonance at muon capture differs from the photo--nuclear
giant resonance by a greater importance of spin waves (similar to 
collective excitations in solids) and by a larger momentum transferred
to the nucleus (neutrino momentum) for muon capture than for photon
absorption with an energy in the vicinity of the giant resonance.
In addition, high--multipolarity transitions play a more significant
part in muon capture than in photo--nuclear reactions.
It is not yet clear why the charged particle yield at muon capture
increases as one goes from $1p-$shell nuclei to $(2s-1d)-$shell
nuclei.
Structure peculiarities of the giant resonance in $(2s-1d)-$shell
nuclei~\cite{goula82,doixx90,eramz77,kozlo74} may play an important
role, though.

For example, the entrance states of one particle--one hole $(1p-1h)$
nuclei should quickly decay into more complicated configurations which
may emit various particles before a thermodynamic equilibrium is
established in the nucleus. 
This is the so-called decay from the pre--equilibrium
state~\cite{balas98,kozlo74}.
In accordance with it, energy spectra of emitted protons and deuterons 
from $(2p-2h)$ states of the daughter nucleus must be well extended
into the high energy region. 
In Ref.~\cite{ubera65} the authors assumed that proton emission at
muon capture may indicate the presence of $(2p-2d)$ states in the
giant dipole configuration.

While in the low--energy region of emitted charged particles the
resonant muon capture mechanism dominates, in the high--energy region
the direct muon capture by correlated nucleon pairs seems to become
prevailing.
In the light of the aforesaid it is interesting to study muon capture 
by ${}^{3}\mathrm{He}$ (and ${}^{4}\mathrm{He}$) nuclei followed
by emission of protons
\begin{equation}
      \label{eq1}
      \mu^{ - }+^{3}\! \mathrm{He} \to \mathrm{p} + \mathrm{n}
      +\mathrm{n} + \nu_{\mu },
\end{equation}
and deuterons
\begin{equation}
      \label{eq2}
      \mu^{ - }+^{3}\! \mathrm{He} \to \mathrm{d} + \mathrm{n} +
      \nu_{\mu} \, .
\end{equation}
Note that muon capture by ${}^{3}\mathrm{He}$ is predominantly (70\%
of the cases) followed by the emission of tritons
\begin{equation}
      \label{eq2a}
      \mu^{ - }+^{3}\! \mathrm{He} \to \mathrm{t} + \nu_{\mu } \, .
\end{equation}
However, this reaction was not studied in our experiment.

Reactions~(\ref{eq1}) and~(\ref{eq2}) also attract interest because
they are background reactions for the nuclear fusion process in the
$\mathrm{d}\mu {}^{3}\mathrm{He}$ molecule
\begin{equation}
      \label{eq3}
      \mathrm{d}\mu^{3}\mathrm{He} \to \mathrm{p} + \alpha + \mu,
\end{equation}
to which considerable
experimental~\cite{mulha98prop,maevx99,delro99,balin98,borei98,knowl01}
and theoretical~\cite{penko97,czapl96,bogda97,bogda99,bystr99b}
studies have been devoted in the past five years.
In addition,  the study of such systems gives the possibility
of verifying fundamental symmetries in strong interactions (such as
charge symmetry or isotope invariance)~\cite{merku87,friar87} and
solving some astrophysical problems~\cite{belya95}.

%The actual knowledge of processes~(\ref{eq1}) and~(\ref{eq2}) is given
%below.
% 
There has been only one experiment~\cite{kuhnx94,cummi92} in which
differential probabilities for muon capture by ${}^{3}\mathrm{He}$
nuclei with the production of protons $ d\lambda _\mathrm{cap}^p /
dE_p $ and deuterons $d\lambda _\mathrm{cap}^d / dE_d $ were measured
at a few proton energies $E_{p}$ in the range $17 - 52$~MeV and
deuteron energies $E_{d}$ in the range $20 - 28$~MeV\@.
In addition, total summed rates for processes shown in
Eqs.~(\ref{eq1}) and~(\ref{eq2}) were measured in three
experiments~\cite{zaimi63b,auerb65,maevx96} and calculated in
Refs.~\cite{yanox64,phili75,congl94}.
% 
%As seen, the experimental and calculated data that could be found in
%the literature were rather limited at that time.
%
A recent review~\cite{measd01} is devoted to the experimental and
theoretical study of the nuclear muon capture and in particular to
the muon capture by He nuclei.
It contains essentially the full list of theoretical and experimental
work performed in this field until today.

%It is worth mentioning a few 
Other points indicating the importance and the necessity of studying
processes of muon capture by ${}^{3}\mathrm{He}$ nuclei are the following:
\begin{itemize}
\item[--] Progress in the wave function calculations for the initial
          and final states of such a three--body
          system~\cite{glock90,skibi99,nogga97,glock96,congl93,congl96}
          will give a better comparison between experiment and theory.
\item[--] Precise information on the characteristics of
          reactions~(\ref{eq1}) and~(\ref{eq2}) in a ``softer'' proton
          and deuteron energy region as that in~\cite{kuhnx94,cummi92}
          by using different techniques will be obtained.
\end{itemize}

The purpose of the study described in this paper is to measure the
energy distributions of protons and deuterons ($S(E_p),\ S(E_d)$)
produced in reactions~(\ref{eq1}) and~(\ref{eq2}).
We will also study the energy dependence of the differential
probabilities for muon capture by ${}^{3}\mathrm{He}$ nuclei.

\section{Experiment}
\label{sec:experiment}

\subsection{Experimental set--up}
\label{sec:experimental-set-up}

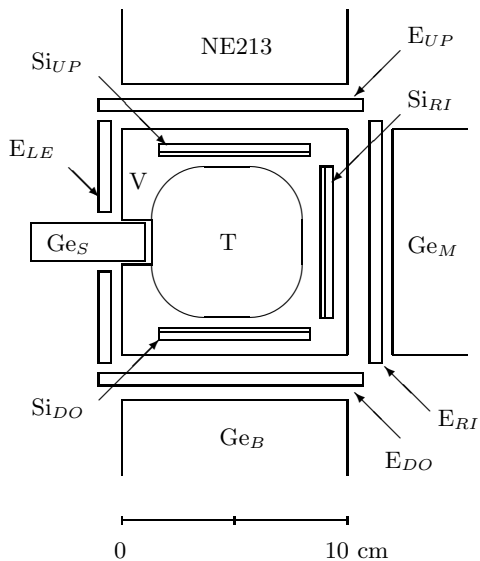
\begin{figure}[t]
  \centerline{
\setlength{\unitlength}{0.1mm}
\begin{picture}(800,800)(0,0)
%\put(20,0){\framebox(780,800){~}}
% place the target
\put(390,450){\oval(200,200)}
\put(380,440){T}
% place the vacuum chamber
%\put(250,300){\framebox(300,300)}
\put(250,300){\line(1,0){300}}
\put(250,600){\line(1,0){300}}
\put(550,600){\line(0,-1){300}}
\put(250,600){\line(0,-1){120}}
\put(250,300){\line(0,1){120}}
\put(250,420){\line(1,0){40}}
\put(250,480){\line(1,0){40}}
\put(290,480){\line(0,-1){60}}
%\put(180,610){\vector(3,-2){115}}
\put(260,520){V}
% place the top of the Ge
\put(130,425){\framebox(150,50)}
\put(150,438){\small Ge$_S$}
% place Si_up, the upper Si detector
\put(300,565){\framebox(200,15)}
\put(300,570){\line(1,0){200}}
\put(130,680){\small Si$_{UP}$}
\put(220,670){\vector(1,-1){90}}
% place Si_do, the lower si detector
\put(300,320){\framebox(200,15)}
\put(300,330){\line(1,0){200}}
\put(130,220){\small Si$_{DO}$}
\put(220,240){\vector(1,1){80}}
% place Si_ri, the right si detector
\put(515,350){\framebox(15,200)}
\put(520,350){\line(0,1){200}}
\put(630,630){\small Si$_{RI}$}
\put(620,620){\vector(-1,-1){90}}
% place E1, the upper electron detector
\put(220,625){\framebox(350,15)}
\put(630,715){\small E$_{UP}$}
\put(620,705){\vector(-1,-1){60}}
% place E2, the right electron detector
\put(580,290){\framebox(15,320)}
\put(670,205){\small E$_{RI}$}
\put(660,225){\vector(-1,1){60}}
% place E3, the lower electron detector
\put(220,260){\framebox(350,15)}
\put(600,150){\small E$_{DO}$}
\put(620,190){\vector(-1,1){60}}
% place E4, the left electron detector
\put(220,290){\framebox(15,120)}
\put(220,490){\framebox(15,120)}
%\dottedline{2}(220,410)(220,490)
%\dottedline{2}(235,410)(235,490)
\put(100,565){\small E$_{LE}$}
\put(180,550){\vector(1,-1){40}}
% place the top NE213
\put(250,660){\line(1,0){300}}
\put(250,660){\line(0,1){100}}
\put(550,660){\line(0,1){100}}
\put(355,700){\small NE213}
% place the medium Germanium (beam right)
\put(610,300){\line(1,0){100}}
\put(610,600){\line(1,0){100}}
\put(610,600){\line(0,-1){300}}
\put(630,435){\small Ge$_M$}
% place the big Germanium
\put(250,240){\line(1,0){300}}
\put(250,240){\line(0,-1){100}}
\put(550,240){\line(0,-1){100}}
\put(380,180){\small Ge$_B$}
% place a scale
\put(250,80){\line(1,0){300}}
\put(250,75){\line(0,1){10}}
\put(240,30){0}
\put(400,75){\line(0,1){10}}
\put(550,75){\line(0,1){10}}
\put(520,30){10 cm}
\end{picture}
}
%  \centerline{\input{app_fut.tex}}
  \caption{Apparatus used in the $\mu$E4 area.  The view is that of
    the incoming muon.  Note that the T1 and T0 scintillators are not
    shown.  The labels are explained in the text.}
  \label{fig:apparatus}
\end{figure}

The experiment was carried out at the $\mu$E4 channel at the Paul
Scherrer Institute (PSI) in Switzerland.
The apparatus was originally designed and used to measure the nuclear
fusion rate in the molecular system $\mathrm{d}\mu
{}^{3}\mathrm{He}$~\cite{mulha98prop,delro99,borei98,knowl01}.
Figure~\ref{fig:apparatus} schematically displays the apparatus as
seen by an incoming muon.

The cryogenic gas target, described in detail in~\cite{borei98},
consisted of a vacuum isolation region (``V'' in
Fig.~\ref{fig:apparatus}) and a~cooled pressure vessel made of pure
aluminum (``T'' in Fig.~\ref{fig:apparatus}).
The pressure vessel enclosed a 66~mm diameter space which was filled
with either pure ${}^{3}\mathrm{He}$ or $\mathrm{D}_2 +
{}^{3}\mathrm{He}$ mixtures.
Five stainless steel flanges held kapton windows over ports in the
pressure vessel to allow the muons to enter and the particles of
interest to escape from the central reaction region.
In total, the target gas volume was $\approx 250\:\mbox{cm}^{3}$.

The incident muons, $\sim17\times10^3\:\mu/$s at momenta 34~MeV/c or
38~MeV/c were detected by a 0.5~mm thick plastic scintillator of area
$45\times45\:\mbox{mm}^2$, called T1, located at the entrance of the
chamber.
The electron impurities in the muon beam were suppressed by a detector
and a lead moderator, called T0, both having aligned $\phi=44$~mm
holes, slightly smaller than T1.
Detectors T0 and T1 are not shown in Fig.~\ref{fig:apparatus} since
they lie in the plane of the paper.
To reduce background coming from muons stopping in the entrance flange
with their subsequent nuclear capture and production of charged
components (protons, deuterons, etc.), a 1~mm thick gold ring was
inserted in the flange hole.
Since the muon lifetime in gold is much shorter than in iron ($\tau
_{Au} \approx 0.073$~$\mu $s, $\tau _{Fe} \approx 0.2$~$\mu
$s~\cite{suzuk87}), the time cut used during the analysis of the
detected event substantially suppresses the background arising from
muon capture by the target body.

Charged muon--capture products were detected by three silicon
telescopes located directly in front of the kapton windows but still
within the cooled vacuum environment (Si$_{UP}$, Si$_{RI}$, and
Si$_{DO}$ in Fig.~\ref{fig:apparatus}).
Each telescope consisted of two Si detectors: a 360~$\mu$m thick
$dE/dx$ detector followed by a 4~mm thick $E$ detector.
The silicon detector preamplifiers and amplifiers were RAL 108--A and
109, respectively~\cite{CLRC}.
Low--energy x~rays from the muon cascade were detected by a
0.17~cm$^{3}$ germanium detector (Ge$_S$ in Fig.~\ref{fig:apparatus})
positioned outside the vacuum chamber, but separated only by several
kapton windows from the reaction volume.
Muon decay electrons were detected by four pairs of plastic
scintillator counters (E$_{LE}$, E$_{UP}$, E$_{RI}$, E$_{DO}$ in
Fig.~\ref{fig:apparatus}) placed around the target.

The gas purity in the target was monitored by 75~cm$^3$ and 122~cm$^3$
germanium detectors (Ge$_M$ and Ge$_{B}$), which were sensitive
to x~rays between 100~keV and 8~MeV\@.
Ge$_M$ and Ge$_B$ were also used to monitor ``harder'' x~rays,
providing information about muon stops in the target walls.
The NE213 detector was used to detect 2.5~MeV neutrons from dd fusion.

The detector electronics triggering system was similar to that used in
experiments performed at TRIUMF (Vancouver, Canada) and details are
given in Ref.~\cite{knowl97}.
The system measured events muon by muon, opening an 8~$\mu$s gate for
each received T1 pulse.
At the end of the event gate, the individual detector electronics were
checked and if any one detector triggered, all detectors were read and
the data stored.
If a second T1 signal arrived during the event gate, we assumed it was
a second muon and discarded the event as pile--up.
Great care was taken with the T1 threshold such that no muons would be
missed, although this increased the rate of event gates started by
electrons.
Those events were rejected in software based on a lower-limit energy
cut from the T1 scintillator.
The pile--up rejection system was much improved over the TRIUMF version
and reduced the detection dead-time for multiple muons from $\approx
50$~ns down to 3~ns.
Thus we had only a $54\times 10^{-6}$ chance per event to have two
muons enter the target simultaneously without our awareness, although
again an upper-limit cut on the T1 energy reduced the number of these
events accepted in the analysis.

\subsection{Experimental conditions}
\label{sec:exper-cond}

\begin{table}[hb]
\begin{ruledtabular}
  \caption{Experimental conditions.  The last column, $N_\mu$, 
    represents the number of muons stopped either in pure
    $^{3}\mathrm{He}$ or in the $\mathrm{D}_2 +
    {}^{3}\mathrm{He}$ mixture.}
  \label{tab:exp-cond}
  \begin{tabular}{ccccccc}
Run & Target & Temp.  & Pressure & $\phi$ &
$\mathrm{c}_{{}^3\mathrm{He}}$ & $N_{\mu}$ \\
&& [K] & [atm] &[LHD] & [\%] & [10$^{6}$] \\ \hline
&&&&&&\\ 
&&& 6.92 &0.0363(7) && \\
&&& 6.85 &0.0359(7) && \\ 
I &${}^{3}\mathrm{He}$ &33 &&&100 &1555.5 \\
&&& 6.78 &0.0355(7) && \\ 
&&& 6.43 &0.0337(7) && \\ \hline 
II &$\mathrm{D}_2 + {}^{3}\mathrm{He}$ &32.8 &5.05  &0.0585(12)&4.96(10)
&4215.6 \\
III &$\mathrm{D}_2 + {}^{3}\mathrm{He}$ &34.5 &12.04 &0.168(12) &4.96(10)
&2615.4 \\
\end{tabular}
\end{ruledtabular}
\end{table}

The experiment was performed using three different gas conditions
which are presented in Table~\ref{tab:exp-cond}.
The first measurement, Run~I, was performed with a pure
${}^{3}\mathrm{He}$ gas at different pressures.
The second and third measurements used a
$\mathrm{D}_{2}+{}^{3}\mathrm{He}$ mixture at two different pressures.
Run~II was performed at 5~atm, whereas Run~III took place at a pressure
more than twice larger, namely 12~atm, where it was necessary to raise
the temperature to avoid liquefying the mixture.
The density $\phi$ is given relative to the standard liquid hydrogen
atomic density (LHD), $N_0 = 4.25\times10^{22}\:\mbox{cm}^{-3}$.
As seen from the last column of Table~\ref{tab:exp-cond}, Run~II was
by far the longest run because its original purpose was to measure the
fusion rate in the $\mathrm{d}\mu {}^{ 3} \mathrm{He}$ molecule and
the muon transfer rate from $\mathrm{d}\mu $ atoms to
${}^{3}\mathrm{He}$ nuclei~\cite{knowl01}.

\section{Measurement method}
\label{sec:measurement-method}

This section describes the method used to measure the differential
muon capture rates by ${}^{3}\mathrm{He}$ nuclei with the production
of protons and deuterons, as given in Eqs.~(\ref{eq1})
and~(\ref{eq2}).
Essentially, it is a simultaneous analysis of the time and energy
spectra of events detected by the Si($dE-E$) counters when muons stop
in the gas target.

The first step is to obtain time and energy spectra from the three
Si($dE-E$) detectors for each run.
As a function of time, we then create two dimensional energy spectra
($dE-E$) to suppress essentially the accidental coincidence background
and to separate precisely the two regions corresponding to the protons
and deuterons.

The second step is to simulate via Monte Carlo (MC) the time and
energy distribution of the events detected by the Si($dE-E$)
detectors.
The simulations are performed as a function of different proton and
deuteron energy distributions.

The final step is a comparison between the experimental results and the MC
simulation.
% 
%The comparison will take place via two different methods.
% 
The first comparison is done using the least--squares analysis between MC
and data, and is described in Sec.~\ref{sec:method-i:-least}.
The second comparison requires first to transform the experimental
spectra such that one 
obtains the initial energy distribution using Bayes theorem.
This analysis is given in Sec.~\ref{sec:method-ii:-bayes}.

The number of protons with a full kinetically allowed energy range 
$\Delta E^\mathrm{max}_{p} = [0; E^\mathrm{max}_{p}]$ 
%in the time interval $[0;t]$ 
per unit of time is, for the case of pure ${}^{3}\mathrm{He}$
\begin{equation}
      \label{eq4}
      \frac{dN_p (\Delta E^\mathrm{max}_{p},t)}{dt} =
      N_{\mu}^{\mathrm{He}} \cdot \lambda _\mathrm{cap}^p \cdot e^{ -
      \lambda_{\mathrm{He}} \cdot t}\, ,
\end{equation}
where $N_{\mu}^{\mathrm{He}}$ is the number of muons
stopping in ${}^{3}\mathrm{He}$ and $\lambda _\mathrm{cap}^p$ is the muon
capture rate in ${}^{3}\mathrm{He}$ when producing a proton.
We use the rate $\lambda_{\mathrm {He}}$ as the sum
\begin{equation}
      \label{eq4b}
      \lambda_{\mathrm {He}} = \lambda _0 +
      \lambda_\mathrm{cap}^{\mathrm{He}} \, 
\end{equation}
where $\lambda _0$ is the free muon decay rate ($\lambda_{0} = 0.4552
\times 10^{6}$~s$^{ - 1}$), and $\lambda_\mathrm{cap}^{\mathrm{He}}$
is the total muon capture rate in ${}^{3}\mathrm{He}$, given by
 \begin{equation}
      \label{eq4a}
      \lambda_\mathrm{cap}^{\mathrm{He}} = \lambda_\mathrm{cap}^p +
      \lambda_\mathrm{cap}^d + \lambda_\mathrm{cap}^t \, .
 \end{equation}
$\lambda_\mathrm{cap}^p$, $\lambda_\mathrm{cap}^d$, and
$\lambda_\mathrm{cap}^t$ are the ${}^{3}\mathrm{He}$ total muon capture
rates when producing a proton, Eq.~(\ref{eq1}), a deuteron,
Eq.~(\ref{eq2}), and a triton, Eq.~(\ref{eq2a}), respectively.
An analogous equation like Eq.~(\ref{eq4}) should also be written for
the production of deuterons.
To avoid complication, we only write equations for the protons using
the $p$ index.

Thus the proton yield produced in the reaction~(\ref{eq1}) during a
time interval $\Delta T = [t_1; t_2]$ for the full energy range $\Delta
E^\mathrm{max}_{p}$ is
\begin{equation}
      \label{eq5}
      N_p (\Delta E^\mathrm{max}_{p}, \Delta T) =
      \frac{N_{\mu}^{\mathrm{He}} \cdot \lambda _\mathrm{cap}^p}
      {\lambda_{\mathrm{He}} } \cdot f_t,
\end{equation}
with the time factor $f_t$ given as
\begin{equation}
      \label{eq6}
      f_t = e^{ - \lambda_{\mathrm{He}} \cdot t_1 } \cdot \left( 1 -
      e^{ - \lambda_{\mathrm{He}} \cdot \delta t}\right),
\end{equation}
where $\delta t = t_2 - t_1$ (here and later in the paper we denote by
$\Delta x = [x_1; x_2]$ the interval of the quantity $x$ and by
$\delta x = x_2 - x_1$ the interval width). 

We are now interested to know the proton yield for a certain energy
range $\Delta E_p = [E_p;E_p + \delta E_p]$ (the proton energy lies
between $E_p$ and $E_p + \delta E_p$).
Such a yield, $N_p (\Delta E_p ,\Delta T)$ is then
\begin{eqnarray}
      \label{eq7}
      N_p (\Delta E_p ,\Delta T) & = & \frac{N_{\mu}^{\mathrm{He}} }
      {\lambda_{\mathrm{He}} } \cdot f_t \int\limits_{E_p}^{E_p +
      \delta E_p} {\frac{d\lambda _\mathrm{cap}^p} {dE_p}dE_p }
      \nonumber \\
      & = & N_{\mu}^{\mathrm{He}} \frac{\lambda
      _\mathrm{cap}^p (\Delta E_p)} {\lambda_{\mathrm{He}} } \cdot f_t
\end{eqnarray}
if one defines
\begin{equation}
      \label{eq7a}
      \lambda _\mathrm{cap}^p (\Delta E_p ) = \int\limits_{E_p}^{E_p +
      \delta E_p} {\frac{d\lambda _\mathrm{cap}^p} {dE_p}dE_p } \, .
\end{equation}

By using Eq.~(\ref{eq7}), one can write the capture rate as function
of the energy range as:
\begin{eqnarray}
      \label{eq9}
      \lambda _\mathrm{cap}^p (\Delta E_p ) 
       & = & \frac{N_p (\Delta E_p ,\Delta T) \cdot
      \lambda_{\mathrm{He}}}{N_{\mu}^{\mathrm{He}} \cdot
      f_t} \nonumber \\
      & = & \frac{N_p (\Delta E_p ,\Delta T^\mathrm{max}) \cdot
      \lambda_{\mathrm{He}}}{N_{\mu}^{\mathrm{He}}}\, ,
\end{eqnarray}
where $\Delta T^\mathrm{max} = [0;\infty)$. 
Therefore the differential capture rate averaged over the proton
energy range becomes
\begin{equation}
      \label{eq8}
      <\frac{d\lambda _\mathrm{cap}^p (E_p)}{dE_p} >= \frac{N_p(\Delta
      {E_p},\Delta T)} {\delta {E_p} }\frac{\lambda_{\mathrm{He}}}
      {N_{\mu}^{\mathrm{He}} }\frac{1}{f_t } \, .
\end{equation}

The number of muon stops in helium $N_{\mu}^{\mathrm{He}} $ is found
by measuring the yield and time distribution of muon decay electrons
stopped in the target (gas and wall).
The total number of muon stops is given by
\begin{equation}
      \label{eq10}
      N_{\mu} = N_{\mu}^{\mathrm{He}} +
      N_{\mu}^{\mathrm{wall}}\, .
\end{equation}
The muon decay electron time spectra can be reproduced by a sum of exponential
functions due to the muon stopping in aluminum and gold (target walls)
as well as in the gas.
\begin{equation}
      \label{eq11}
      \frac{dN_e }{dt} = A_{\mathrm {Al}} \cdot e^{ - \lambda
      _{\mathrm {Al}} \cdot t} + A_{\mathrm {Au}} \cdot e^{ - \lambda
      _{\mathrm {Au}} \cdot t} + A_{\mathrm {He}} \cdot e^{ - \lambda
      _{\mathrm {He}} \cdot t} + B,
\end{equation}
where $A_{\mathrm {Al}}$, $A_{\mathrm {Au}}$, and $A_{\mathrm {He}}$
are the normalization amplitudes and  
\begin{eqnarray}
      \label{eq11a}
      \lambda _{\mathrm {Al}} & = & Q_{\mathrm {Al}} \cdot \lambda_0 +
      \lambda_\mathrm{cap}^{\mathrm {Al} }\, , \nonumber \\
      \lambda _{\mathrm {Au}} & = & Q_{\mathrm {Au}} \cdot \lambda_0 +
      \lambda_\mathrm{cap}^{\mathrm {Au} }\, , \\
      \lambda _{\mathrm {He}} & = & \lambda_0 +
      \lambda_\mathrm{cap}^{\mathrm {He} }\, , \nonumber
\end{eqnarray}
are the muon disappearance rates in the different elements (the rates
are the inverse of muon lifetimes in the target wall materials).
In reality, Eq.~(\ref{eq11}) is an approximation of a more complex
equation, which can be found in Ref.~\cite{knowl99b}.
The nuclear capture rates in aluminum and gold, 
$\lambda_\mathrm{cap}^{\mathrm {Al}}= 0.7054 (13) \times
10^6$~s$^{-1}$ and $\lambda_\mathrm{cap}^{\mathrm {Au}}= 13.07 (28)
\times 10^6$~s$^{-1}$, are taken from~\cite{suzuk87}.
$Q_{\mathrm{Al}}$ and $Q_{\mathrm{Au}}$ are the Huff factors, which
take into account that muons are bound in the $1s$ state of the
respective nuclei when they decay.
This factor is negligible for helium but necessary for aluminum
$Q_{\mathrm {Al}}=0.993$ and important for gold
$Q_{\mathrm{Au}}=0.850$~\cite{suzuk87}.
The constant B characterizes the random coincidence background.

By measuring the amplitude $A_{\mathrm {He}}$
\begin{equation}
      \label{eq12}
      A_{\mathrm {He}} = N_{\mu}^{\mathrm{He}} \cdot
      \varepsilon _e \cdot \lambda _0
\end{equation}
and knowing the electron detection efficiencies $\varepsilon _e$
averaged over the energy distributions, one obtains the number of
muons stopping in helium $N_{\mu}^{\mathrm{He}}$.
The muon decay electron detection efficiency $\varepsilon_{e}$ is
determined experimentally as the ratio 
\begin{equation}
      \label{eq13}
      \varepsilon _e = \frac{N_{x - e} }{N_x },
\end{equation}
where$ N_{x - e}$ is the number of x~rays of the $\mu
{}^{3}\mathrm{He}$ K$\alpha$ line, measured by the germanium detector
(Ge$_S$), in coincidence with a muon decay electron. 
$N_{x}$ is the same number of x~rays of the $\mu {}^{3}\mathrm{He}$
K$\alpha$ line when no coincidence is required.

By determining the quantities $N_p(\Delta E_p,\Delta T)$ and $N_d
(\Delta E_d,\Delta T)$ based on the analysis of the two--dimensional
energy distributions ($dE-E$), knowing
$\lambda_\mathrm{cap}^{\mathrm{He}} = 2216 \,(70) \, \mbox{s}^{ - 1}$
as determined in Ref.~\cite{maevx96} (this value is in good agreement
with the calculated value $\lambda_\mathrm{cap}^{\mathrm{He}} = 2140
\, \mbox{s}^{-1}$ from~\cite{congl94}), we can obtain the muon capture
rate for protons $\lambda _\mathrm{cap}^p (\Delta E_p )$ and deuterons
$\lambda _\mathrm{cap}^d (\Delta E_d )$ as well as both differential
rates $d\lambda _\mathrm{cap}^p/dE_p$ and $d\lambda
_\mathrm{cap}^d/dE_d$ as a function of the proton (deuteron) energy.

\section{The analysis of experimental data}
\label{sec:analys-exper-data}

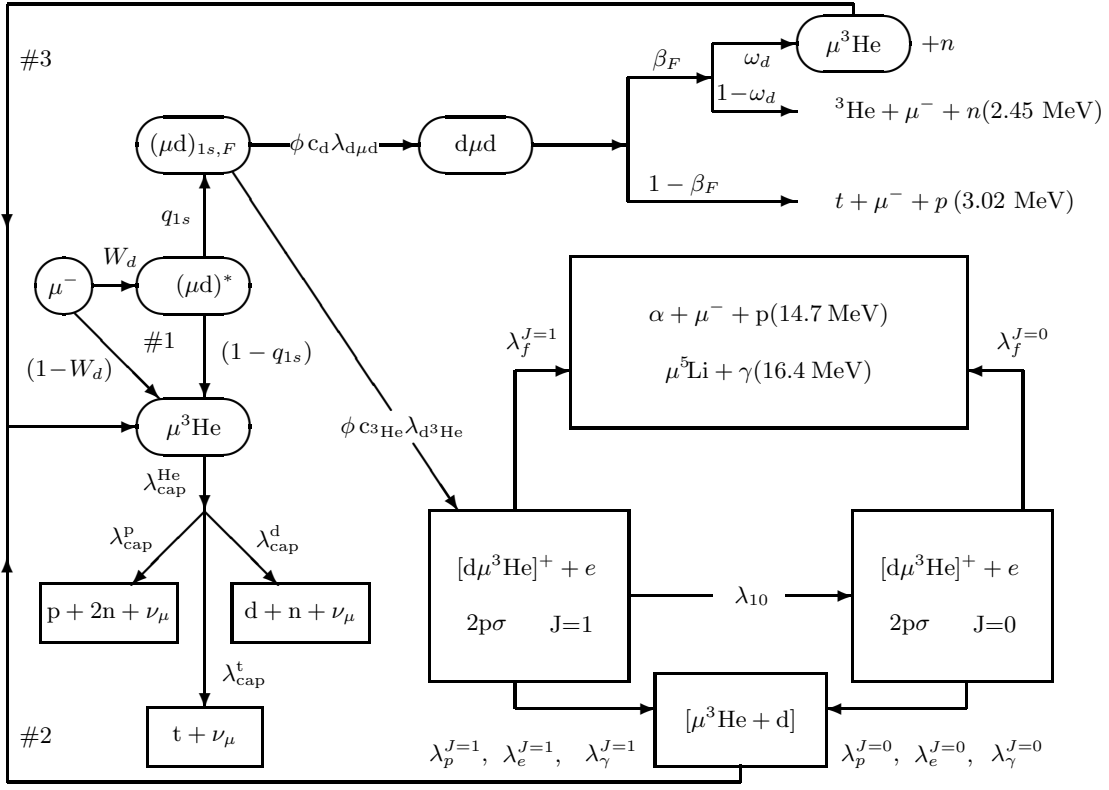
\begin{figure*}[t]
\centerline{
\setlength{\unitlength}{0.75mm}
\thicklines
\begin{picture}(200,145)(0,-10)
%\put(-10,-13){\framebox(210,156){~}}
%  place the mu
\put(10,85){\oval(10,10)}
\put(10,85){\makebox(0,0){$\mu^{-}$}}
%place the lines for the rates
% place the rates
\put(15,85){\vector(1,0){8}}
\put(20,90){\makebox(0,0){$W_{d}$}}
\put(12,80){\vector(1,-1){15}}
\put(11,70){\makebox(0,0){$(1\!-\!W_{d})$}}
%with a number 1
\put(27,75){\makebox(0,0){\#1}}
% q1s
\put(35,90){\vector(0,1){15}}
\put(30,97){\makebox(0,0){$q_{1s}$}}
% 1-q1s
\put(35,80){\vector(0,-1){15}}
\put(46,73){\makebox(0,0){$(1-q_{1s})$}}
%    place the dmu excited
\put(33,85){\oval(20,10)}
\put(35,85){\makebox(0,0){$(\mu {\mathrm d} )^{*}$}}
% place the dmu ground
\put(33,110){\oval(20,10)}
\put(33,110){\makebox(0,0){$(\mu {\mathrm d})_{1s,F}$}}
%  place the helium
\put(33,60){\oval(20,10)}
\put(33,60){\makebox(0,0){$\mu{}^3\mathrm{He} $}}
%place the capture rate
\put(35,55){\vector(0,-1){10}}
\put(28,50){\makebox(0,0){$\lambda_\mathrm{cap}^{\mathrm{He}}$}}
% all about capture in helium
\put(6,22){\framebox(24,10){}}
\put(18,27){\makebox(0,0){$\mathrm{p} + 2 \mathrm{n} + \nu_\mu$}}
\put(40,22){\framebox(24,10){}}
\put(52,27){\makebox(0,0){$\mathrm{d} + \mathrm{n} + \nu_\mu$}}
\put(25,0){\framebox(20,10){}}
\put(35,5){\makebox(0,0){$\mathrm{t} + \nu_\mu$}}
%vector and rate for capture in helium
\put(35,45){\vector(1,-1){13}}
\put(48,40){\makebox(0,0){$\lambda_\mathrm{cap}^{\mathrm{d}}$}}
\put(35,45){\vector(-1,-1){13}}
\put(22,40){\makebox(0,0){$\lambda_\mathrm{cap}^{\mathrm{p}}$}}
\put(35,45){\vector(0,-1){35}}
\put(42,16){\makebox(0,0){$\lambda_\mathrm{cap}^{\mathrm{t}}$}}
%place the ddmu formation lines and rates
\put(43,110){\line(1,0){7}}
\put(58,110){\makebox(0,0){$\phi \, \mathrm {\mathrm c}_{\mathrm d}
\lambda_{\mathrm{d}\mu\mathrm{d}}$}}
\put(66,110){\vector(1,0){7}}
%place the dmuhe formation lines and rates
\put(40,105){\line(2,-3){28}}
\put(70,60){\makebox(0,0){$\phi \, \mathrm {\mathrm c}_{{}^3\mathrm{He}} \lambda_{{\mathrm d}{}^3\mathrm{He}}$}}
\put(72,57){\vector(2,-3){8}}
% place the He-dmu excited boxes and molecules   J=1
\put(75,15){\framebox(35,30){}}
\put(92,35){\makebox(0,0){$[\mathrm{d}\mu{}^3\mathrm{He}]^+ + e$}}
\put(85,25){\makebox(0,0){2p$\sigma$}}
\put(100,25){\makebox(0,0){J=1}}
% place the He-dmu excited boxes and molecules   J=0
\put(150,15){\framebox(35,30){}}
\put(167,35){\makebox(0,0){$[\mathrm{d}\mu{}^3\mathrm{He}]^+ + e$}}
\put(160,25){\makebox(0,0){2p$\sigma$}}
\put(175,25){\makebox(0,0){J=0}}
% place the line for passage J=1 to J=0
\put(110,30){\line(1,0){15}}
\put(132,30){\makebox(0,0){$\lambda_{10}$}}
\put(138,30){\vector(1,0){12}}
% place the He-dmu fusion boxes and results
\put(100,60){\framebox(70,30){}}
\put(135,80){\makebox(0,0){$\alpha + \mu ^{-} + {\mathrm p}(14.7\:\mathrm{MeV})$}}
\put(135,70){\makebox(0,0){$\mu{}^{5}\!\mathrm{Li}+\gamma(16.4\:\mathrm{MeV})$}}
% place the He-dmu ground boxes and molecules  
\put(115,0){\framebox(30,16){}}
\put(130,8){\makebox(0,0){$[\mu {}^3\mathrm{He} + {\mathrm d}]$}}
%place the lines for ground
\put(90,15){\line(0,-1){5}}
\put(90,10){\vector(1,0){25}}
\put(170,15){\line(0,-1){5}}
\put(170,10){\vector(-1,0){25}}
%place the rates for ground
\put(80,2){\makebox(0,0){$\lambda^{J=1}_{p}$,}}
\put(93,2){\makebox(0,0){$\lambda^{J=1}_{e}$,}}
\put(107,2){\makebox(0,0){$\lambda^{J=1}_{\gamma}$}}
\put(153,2){\makebox(0,0){$\lambda^{J=0}_{p}$,}}
\put(166,2){\makebox(0,0){$\lambda^{J=0}_{e}$,}}
\put(179,2){\makebox(0,0){$\lambda^{J=0}_{\gamma}$}}
%place the lines for fusion
\put(90,45){\line(0,1){25}}
\put(90,70){\vector(1,0){10}}
\put(180,45){\line(0,1){25}}
\put(180,70){\vector(-1,0){10}}
%place the rates for fusion
\put(93,75){\makebox(0,0){$\lambda^{J=1}_f$}}
\put(180,75){\makebox(0,0){$\lambda^{J=0}_f$}}
% place dd\mu
\put(83,110){\oval(20,10)}
\put(83,110){\makebox(0,0){${\mathrm d} \mu {\mathrm d} $}}
\put(93,110){\vector(1,0){17}}
% start the tree
\put(110,100){\line(0,1){22}}
%place the branching rate
\put(110,122){\vector(1,0){15}}
\put(117,125){\makebox(0,0){$\beta_F$}}
% second tree top
\put(125,116){\line(0,1){12}}
%place the branching rate
\put(125,128){\vector(1,0){15}}
\put(133,125){\makebox(0,0){$\omega_{d}$}}
\put(150,128){\oval(20,10)}
\put(150,128){\makebox(0,0){$\mu {}^3\mathrm{He}$}}
\put(165,128){\makebox(0,0){$+ n$}}
%place the branching rate
\put(125,116){\vector(1,0){15}}
\put(131,119){\makebox(0,0){$1\!-\!\omega_{d}$}}
\put(170,116){\makebox(0,0){$\:{}^3\mathrm{He} + \mu^{-} + n {\mathrm{(2.45~MeV)}}$}}
% first tree bottom
%place the branching rate
\put(110,100){\vector(1,0){30}}
\put(120,103){\makebox(0,0){$1-\beta_F$}}
\put(168,100){\makebox(0,0){$t + \mu^{-} + p\:{\rm (3.02~MeV)}$}}
%place the cycle after dd-fusion
\put(150,133){\line(0,1){2}}
\put(150,135){\line(-1,0){150}}
\put(0,135){\vector(0,-1){40}}
\put(0,95){\line(0,-1){35}}
\put(0,60){\vector(1,0){23}}
%with a number 2
\put(5,125){\makebox(0,0){\#3}}
%place the cycle after d3He-fusion
\put(130,0){\line(0,-1){3}}
\put(130,-3){\line(-1,0){130}}
\put(0,37){\line(0,1){23}}
\put(0,-3){\vector(0,1){40}}
%with a number 3
\put(5,5){\makebox(0,0){\#2}}
\end{picture}
}
%      \centerline{\input{kinetics.tex}}
      \caption{Scheme of muon processes in the $\mathrm{D}_2 +
               {}^{3}\mathrm{He}$ mixture. Muon capture by
               $^{3}\mathrm{He}$ occurs via process \#1 (with
               $\sim$30\% yield). The essential part of the capture
               (approximatively 65\%) occurs after the $\mathrm{d}\mu
               {}^{3}\mathrm{He}$ formation (\#2). A small amount of
               capture is occurring after $\mathrm{d}\mu\mathrm{d}$
               fusion (\#3). Details about all processes and rates are
               found in Ref.~\cite{knowl01}.}
\label{fig:diagram}
\end{figure*}

As already mentioned in Sec.~\ref{sec:exper-cond}, the experiment was
performed using two different gases, namely pure $^{3}\mathrm{He}$ as
well as a mixture of $\mathrm{D}_{2}+{}^{3}\mathrm{He}$.
When a muon is stopped in the gas mixture, different processes occur.
A diagram of processes occurring in the
$\mathrm{D}_{2}+{}^{3}\mathrm{He}$ mixture (the most complex one) is
displayed in Fig.~\ref{fig:diagram}.

In the run with pure $^{3}\mathrm{He}$ (Run~I of
Table~\ref{tab:exp-cond}) the quantities $\lambda
_\mathrm{cap}^{p}(\Delta E_{p} )$ and $\lambda
_\mathrm{cap}^{d}(\Delta E_{d} )$ for the protons and the deuterons
are found according to Eq.~(\ref{eq9}).
In the runs with a $\mathrm{D}_{2}+{}^{3}\mathrm{He}$ mixture (Runs~II
and~III of Table~\ref{tab:exp-cond}) the same rates are found as
follows. The number of protons per time unit 
\begin{eqnarray}
      \label{eq14}
      \frac{dN_{p} }{dt} & = & N_{\mu}^{\mathrm{D/He}}
      \lambda _\mathrm{cap}^{p} \left[ \mathrm{a}_{\mathrm{He}}
      \cdot e^{ -\lambda_{\mathrm {He}} \cdot t} \right. \nonumber \\ 
      & + & \left. \xi_{\mathrm{D}} \cdot
      (e^{ - \lambda_{\mu \mathrm{He}} \cdot t} -
      e^{ - \lambda_{\mu \mathrm{d}} \cdot t})]  , \right. \\
      \xi_{\mathrm{D}} & = &  
      \frac{\mathrm{q}_{1s}\cdot \mathrm{a}_{\mathrm{D}}\cdot
      \lambda _{\mathrm{d}{}^3\mathrm{He}}\cdot \phi\cdot \,
      \mathrm{c}_{{}^3\mathrm{He}} }{\lambda_{\mu \mathrm{d}} -
      \lambda_{\mathrm {He}} } ,  \nonumber 
%(e^{ -\lambda_{\mathrm {He}} \cdot t} -
%      e^{ - \lambda_{\mu \mathrm{d}} \cdot t}) \right] , \nonumber
\end{eqnarray}
with $N_{\mu}^{\mathrm{D/He}} $ the number of muon stops in the
$\mathrm{D}_2 + {}^{3}\mathrm{He}$ mixture and $\lambda
_{\mathrm{d}{}^3\mathrm{He}}\cdot \phi\cdot \,
\mathrm{c}_{{}^3\mathrm{He}}$ the experimental molecular
$\mathrm{d}\mu{}^3\mathrm{He}$ formation rate, using the known value
$\lambda_{\mathrm {d}{}^3\mathrm{He}} = 2.42 \,(18) \times 10^{8} \,
\mbox{s}^{ - 1}$~\cite{knowl01}.
The rate $\lambda_{\mathrm{He}} = 0.457 \times 10^6 \, \mbox{s}^{-1}$
is given by Eq.~(\ref{eq4b}) using the known total capture
rate~\cite{maevx96}.
%,acker98}.
%
$\phi$ and $\mathrm{c}_{{}^3\mathrm{He}}$ are the target density and helium
concentration given in Table~\ref{tab:exp-cond}.  
The experimental disappearance rate $\lambda_{\mu \mathrm{d}}$ 
for the $\mathrm{d}\mu$ atom in the ground state is given as
\begin{equation}
      \label{eq16}
      \lambda_{\mu \mathrm{d}} = \lambda _0 +
      \lambda_{\mathrm{d}\mu\mathrm{d}}\cdot \phi\cdot \, 
       \mathrm {c}_{\mathrm {d}}\cdot
      \tilde {\omega }_d + \lambda_{\mathrm {d}{}^3\mathrm{He}}\cdot 
      \phi\cdot \, \mathrm{c}_{{}^3\mathrm{He}}
\end{equation}
using the $\mathrm{d}\mu\mathrm{d}$ molecular formation rate
$\lambda_{\mathrm{d}\mu\mathrm{d}} = 0.05 \times 10^{6} \, \mbox{s}^{
- 1}$~\cite{petit96,vorob88,gartn00,zmesk90} and the effective muon
sticking coefficient to the ${}^{3}\mathrm{He}$ nucleus resulting form
the nuclear fusion reaction in the $\mathrm{d}\mu \mathrm{d}$ molecule
$\mathrm{d} \mu \mathrm{d} \to \mu {}^{3}\mathrm {He} + n$, $\tilde
{\omega}_d = 0.07$~\cite{vorob88,balin84}.
The deuterium concentration $\mathrm{c}_\mathrm{d} = 1 -
\mathrm{c}_{{}^3\mathrm{He}}$ is obtained from
Table~\ref{tab:exp-cond}.
In reality, Eq.~(\ref{eq16}) is an approximation of a more complex
equation, which can be found in Ref.~\cite{knowl99b}.

The total probability $\mathrm{a}_{\mathrm{He}}$ for $\mu
{}^{3}\mathrm {He}$ formation is
\begin{equation}
      \label{eq15a}
      \mathrm{a}_{\mathrm{He}} = \mathrm{a}_{\mathrm{He}}^0 +
      \mathrm{a}_{\mathrm{He}}^1
\end{equation}
where $\mathrm{a}_{\mathrm{He}}^0 $ is the muon capture probability by
${}^{3}\mathrm{He}$ and $\mathrm{a}_{\mathrm{He}}^1$ is the
probability of muon transfer from an excited state of the
$\mathrm{d}\mu$ atom to ${}^{3}\mathrm{He}$.
Explicitly, 
\begin{eqnarray}
      \label{eq15}
      \mathrm{a}_{\mathrm{He}}^0 & = &\frac{A\cdot 
      \mathrm{c}_{{}^3\mathrm{He}} }{1 + A \cdot
      \mathrm{c}_{{}^3\mathrm{He}}}\, ,  \nonumber \\
      \mathrm{a}_{\mathrm{He}}^1 & = & (1 -
      \mathrm{q}_{1s})\mathrm{a}_\mathrm{D}\, , \\ 
      \mathrm{a}_{\mathrm{D}} & = &\frac{1}{1 + A\cdot 
      \mathrm{c}_{{}^3\mathrm{He}} } \, , \nonumber 
\end{eqnarray}
where $A$ is the ratio between the stopping powers of the deuterium
and helium atoms, $A = 1.7 \, (2)$~\cite{bystr95c} and
$\mathrm{a}_{\mathrm{D}} $ is the muon capture probability by a
deuterium atom.
$\mathrm{q}_{1s}$ is the probability that the excited $(\mathrm{d}\mu )^*$
atom will reach the ground--state.
The term $\mathrm{q}_{1s}\cdot \mathrm{a}_{\mathrm{D}}$ is the probability
for a muon stopped in the $\mathrm{D}_2 + {}^3\mathrm{He}$ mixture to be
captured by a deuterium atom and reach the ground--state.
The $\mathrm{q}_{1s}$ values for the Runs~II and~III are 0.80 and 0.72,
respectively, according to Refs.~\cite{bystr90e,bystr95c,bystr99}.
These values are somewhat higher than a recent
experiment~\cite{augsb03} ($\mathrm{q}_{1s} = 0.689 \, (27)$)
performed at an intermediate ${}^3\mathrm{He}$ concentration
($\mathrm{c}_{{}^3\mathrm{He}} = 9.13 \%$).
Using  Eqs.~(\ref{eq15a}) and~(\ref{eq15}) one can then write
\begin{equation}
      \label{eq15b}
      \mathrm{a}_{\mathrm{He}} =\frac{1}{1 + A \cdot
      \mathrm{c}_{{}^3\mathrm{He}} }(1 - \mathrm{q}_{1s} + A\cdot 
      \mathrm{c}_{{}^3\mathrm{He}} ) \, .
\end{equation}

Thus the proton yield in the time interval $\Delta T = [t_1;t_2]$, for
the whole energy range $\Delta E^\mathrm{max}_{p}$ is given by
\begin{equation}
      \label{eq17}
      N_{p} (\Delta E^\mathrm{max}_{p}, \Delta T) =
      \frac{N_{\mu}^{\mathrm{D/He}} \cdot \lambda
      _\mathrm{cap}^p}{\lambda_{\mathrm{He}} } \cdot f_t,
\end{equation}
with the time factor $f_t$ given as
\begin{eqnarray}
      \label{eq18}
      f_t & = & ( \mathrm{a}_{\mathrm{He}} + \xi_{\mathrm{D}})
      ( e^{ -\lambda_{\mathrm {He}} \cdot t_1 } - e^{ - \lambda_{\mathrm
      {He}} \cdot t_2 })  \nonumber \\
      & - & \xi_{\mathrm{D}} \frac{\lambda_{\mathrm{He}}}
      {\lambda_{\mu \mathrm{d}}}(e^{ - \lambda_{\mu \mathrm{d}} \cdot
      t_1 } - e^{ - \lambda_{\mu \mathrm{d}} \cdot t_2 }) \, . 
\end{eqnarray}
The number of protons following  muon capture in the energy range
$\Delta E_{p}$ is then
\begin{equation}
      \label{eq19}
      N_{p} (\Delta E_{p} ,\Delta T) =
      N_{\mu}^{\mathrm{D/He}} \cdot f_t \cdot
      \frac{\lambda_\mathrm{cap}^{p} (\Delta E_{p} )}{
      \lambda_{\mathrm{He}} }
\end{equation}
and the capture rate becomes
\begin{equation}
      \label{eq20}
      \lambda _\mathrm{cap}^{p} (\Delta E_{p} ) = \frac{N_{p} (\Delta
      E_{p} ,\Delta T) \cdot \lambda_{\mathrm{He}}
      }{N_{\mu}^{\mathrm{D/He}} \cdot f_t }\, .
\end{equation}
Note that Eqs.~(\ref{eq20}) and~(\ref{eq9}) are similar for both the
pure gas and the mixture.
The difference lies in the time factor $f_t$, given by
Eqs.~(\ref{eq18}) and~(\ref{eq6}).

The calculation of $f_t$ for the $\mathrm{D}_2 + {}^{3}\mathrm{He}$
mixture (Eq.~(\ref{eq18})) demands the previous knowledge of
$\mathrm{a}_{\mathrm{He}}$, $\lambda_{\mathrm {d}{}^3\mathrm{He}}$,
$\lambda_{\mathrm{d} \mu \mathrm{d}}$, $\lambda_{\mathrm {He}}$, and
$\lambda_{\mu \mathrm{d}}$.
Even if most of those values are known from other experiments, this
experiment allows us another independent determination of these 
quantities and hence a consistency check.
The rate of $\lambda_{\mu \mathrm{d}}$ is found by analyzing the time
distribution of either the proton, the deuteron, or the photon emitted
after $\mathrm{d}\mu{}^3\mathrm{He}$ formation.
The time distribution can be fitted using Eq.~(\ref{eq14}). 
For Run~I with pure ${}^{3}\mathrm{He}$ the value of $f_t$ was
determined by using $\lambda_\mathrm{cap}^{\mathrm{He}} $ in
Eq.~(\ref{eq4}).

\begin{figure}[t]
      \includegraphics[angle=90,width=0.48\textwidth]{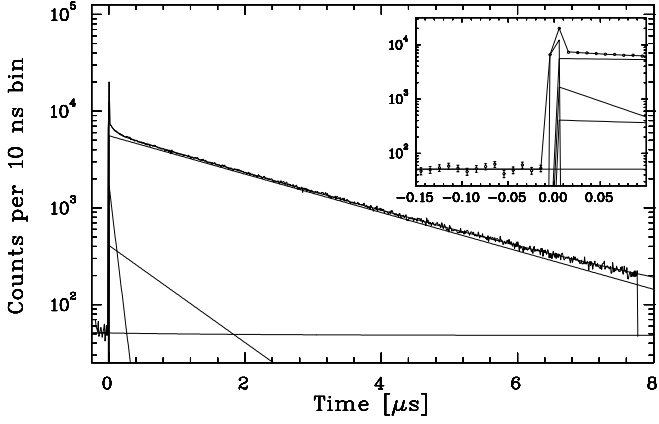}
      \caption{Time distribution of muon decay electrons measured in
               Run~I with pure ${}^{3}\mathrm{He}$.  Top--right
               picture shows details of early times.}
      \label{fig:electron}
\end{figure}

\begin{figure}[b]
      \includegraphics[angle=90,width=0.48\textwidth]{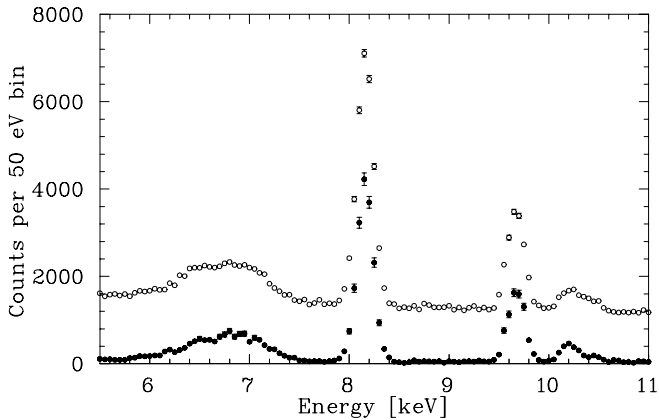}
      \caption{Muonic x--ray spectra measured by the germanium
               detector in a mixture of $\mathrm{D}_2 +
               {}^{3}\mathrm{He}$,  without (open circles) and
               with (solid circles) muon decay electron coincidences.}
      \label{fig:xray}
\end{figure}

As mentioned in Sec.~\ref{sec:measurement-method}, we find the number
of muon stops in the gas by fitting Eq.~(\ref{eq11}) to the muon decay
electron time distributions.
Figure~\ref{fig:electron} represents such a fit of electron time
spectra when all four detector pairs E$_{UP}$, E$_{RI}$, E$_{DO}$, and
E$_{LE}$ are added together.

Figure~\ref{fig:xray} displays the energy spectra of the low--energy
photons from $\mu {}^{3}{\mathrm He}$ atoms (K${\alpha}$ at 8.2~keV,
K${\beta}$ at 9.6~keV, and K${\gamma }$ at 10.2~keV) measured with the
germanium detector Ge$_S$ with and without the delayed muon decay electron
coincidence.
The electron detection efficiency $\varepsilon_e$ is determined using
Eq.~(\ref{eq13}).
The so obtained value still needs to be corrected for the difference
in positions between the germanium and the Si($dE-E$) detectors with
respect to the muon stop distribution along the incident muon beam.
The final value for the total muon decay electron detection efficiency
of the four electron counters found from the analysis of Run~II is
$\varepsilon_e = 16.4 \pm 0.22${\%}~\cite{delro99,knowl01}.

Since the background is mainly caused by muon stops in the target
walls (Al, Au) followed by their nuclear capture and the emission of
charged products (with characteristic times $\tau_{\mathrm {Al}} =
0.865$~$\mu$s and $\tau_{\mathrm {Au}} =
0.073$~$\mu$s~\cite{suzuk87}), the background contribution will be
determined in two steps.

The first step is to remove the background contribution from muon
stops in gold.
Hence, we selected only events detected by the Si($dE-E$) detectors
for times $t > 4 \tau_{\mathrm {Au}}$.
The remaining events are due to muon stops in the gas, which have a
time distribution following Eq.~(\ref{eq4}) for pure $^{3}\mathrm{He}$
and Eq.~(\ref{eq14}) for the mixture $\mathrm{D}_2 +
{}^{3}\mathrm{He}$, and muon stops in aluminum.
Therefore, the time distribution of the Si($dE-E$) events in pure
$^{3}\mathrm{He}$ is represented as:
\begin{equation}
      \label{eq21}
      \frac{dN_{p}^\mathrm{meas} }{dt} = D_1 \cdot e^{ - \lambda
      _{\mathrm {Al}} \cdot t} + D_2 \cdot e^{ - \lambda_{\mathrm
      {He}} \cdot t} + C
\end{equation}
with
\begin{eqnarray}
      \label{eq21a}
      D_1 & = & N_{\mu}^{\mathrm {Al}} \cdot \lambda
      _\mathrm{cap}^{\mathrm {Al}} \cdot \tilde{\varepsilon
      }_{p}^{\mathrm {Al}} \nonumber \\ 
      D_2 & = & N_{\mu}^{\mathrm{He}} \cdot \lambda
      _\mathrm{cap}^{p} (\Delta E_{p} ) \cdot \tilde {\varepsilon
      }_{p}^{\mathrm{He}} \, .
\end{eqnarray}
The terms $N_{\mu}^{\mathrm{He}}$ and $N_{\mu}^{\mathrm{Al}}$
represent the number of muons stopping in helium and aluminum,
respectively, $\tilde {\varepsilon }_{p}^{Al} $ and $\tilde
{\varepsilon }_{p}^{\mathrm{He}} $ are the proton detection
efficiencies after muon capture in aluminum or helium averaged over
the energy interval $\Delta E_{p}$, and $C$ is the accidental
coincidence background.

For the $\mathrm{D}_2 + {}^{3}\mathrm{He}$ mixture, Eq.~(\ref{eq21})
has to be rewritten as
\begin{equation}
      \label{eq21b}
      \frac{dN_{p}^\mathrm{meas} }{dt} = D_1 \cdot e^{ - \lambda
      _{\mathrm {Al}} \cdot t} + D'_2 \cdot e^{ - \lambda_{\mathrm
      {He}} \cdot t} - D'_3 \cdot e^{ - \lambda_{\mu \mathrm{d}} \cdot
      t} + C
\end{equation}
with
\begin{eqnarray}
      \label{eq21c}
      D'_2 & = & N_{\mu}^{\mathrm{D/He}} \cdot \lambda
      _\mathrm{cap}^{p} (\Delta E_{p} ) \cdot \tilde {\varepsilon
      }_{p}^{\mathrm{He}}\cdot (\mathrm{a}_{\mathrm{He}} + 
      \xi_{\mathrm{D}})\ , \nonumber \\
%
%      &  \times & \left[
%      \mathrm{a}_{\mathrm{He}} + \frac {\mathrm{q}_{1s}
%      \mathrm{a}_{\mathrm{D}} { \lambda_ {\mathrm{d}{}^3\mathrm{He}}}
%      \phi \, \mathrm{c}_{{}^3\mathrm{He}}} {\lambda_{\mu \mathrm{d}} -
%      \lambda_{\mathrm {He}}} \right] \nonumber \\
%
      D'_3 & = & N_{\mu}^{\mathrm{D/He}} \cdot \lambda
      _\mathrm{cap}^{p} (\Delta E_{p} ) \cdot \tilde {\varepsilon
      }_{p}^{\mathrm{He}}\cdot \xi_{\mathrm{D}}\ ,
% 
%      & \times & \frac {\mathrm{q}_{1s}
%      \mathrm{a}_{\mathrm{D}} { \lambda_ {\mathrm{d}{}^3\mathrm{He}}}
%      \phi \, \mathrm{c}_{{}^3\mathrm{He}}} {\lambda_{\mu \mathrm{d}} -
%      \lambda_{\mathrm {He}}} \, , \nonumber 
\end{eqnarray}
where the term $N_{\mu}^{\mathrm{D/He}}$ represents the
number of muons stopping in the mixture.
The constant $D_2$ is replaced by the corresponding $D'_2$ and
$D'_3$.

\begin{figure}[t]
      \includegraphics[angle=90,width=0.46\textwidth]{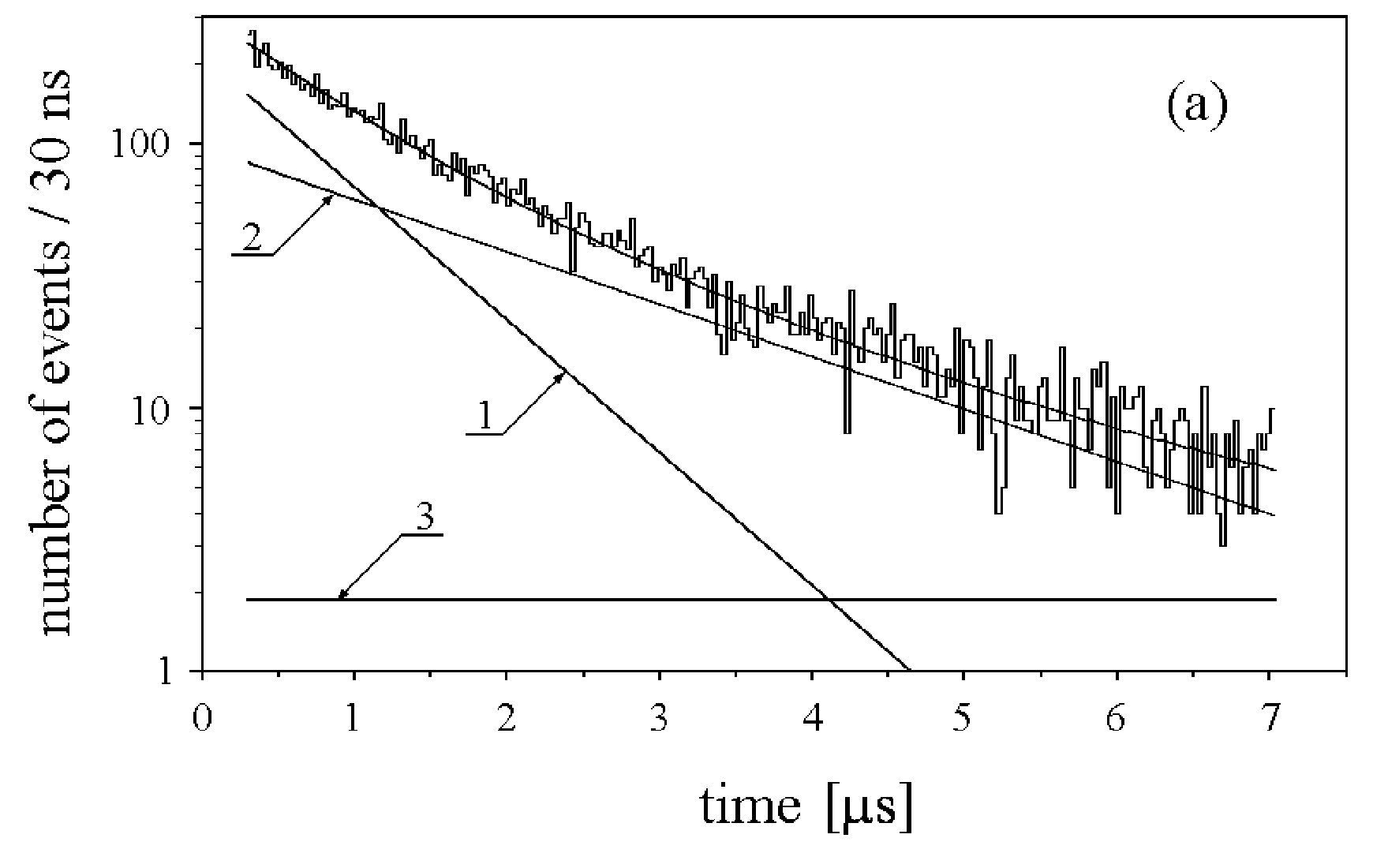}
      \includegraphics[angle=90,width=0.46\textwidth]{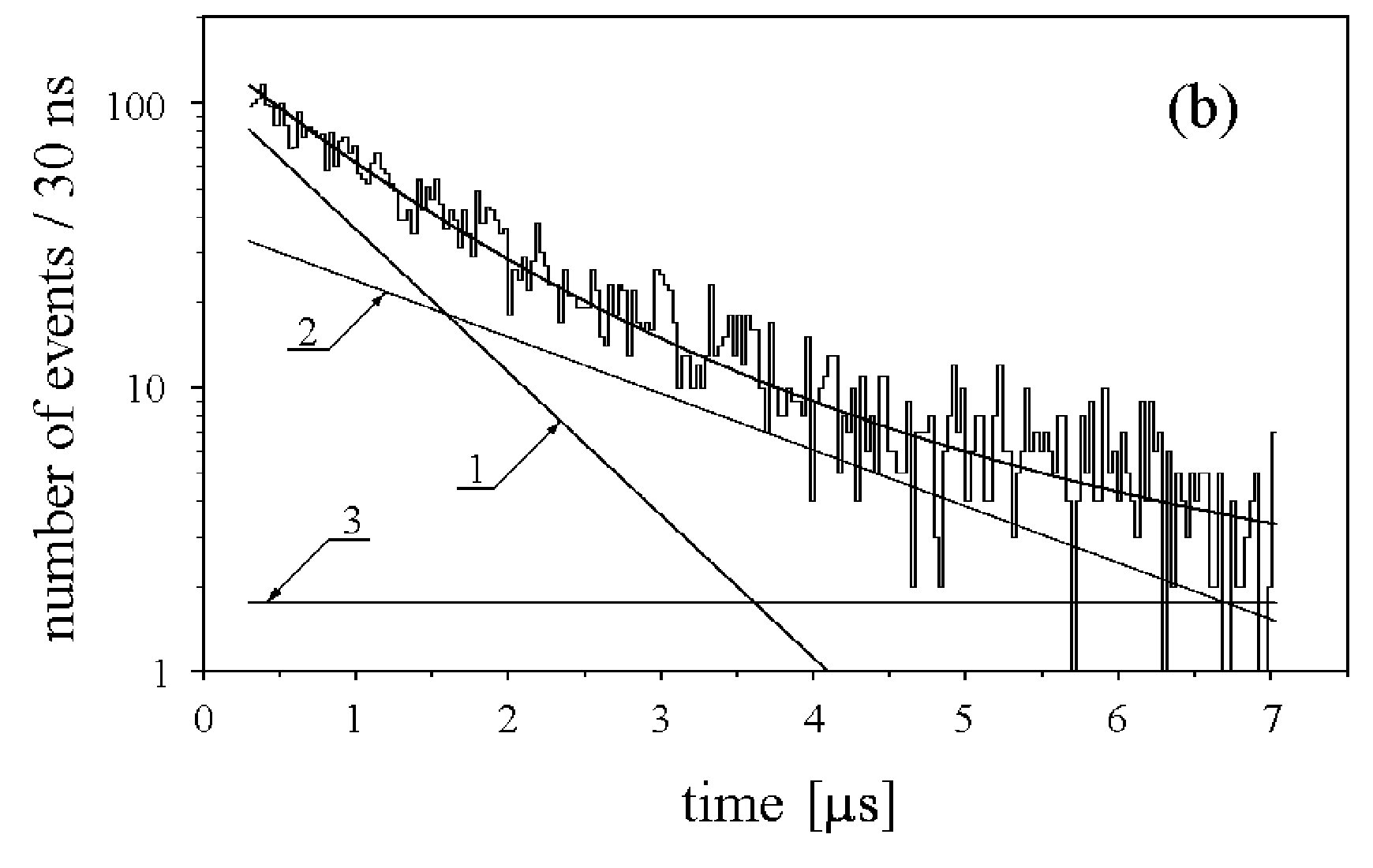}
      \caption{Time distributions of Si($dE-E$) events for Run~I: (a)
               protons, (b) deuterons.  The histograms represent the
               experimental data.  The solid lines 1 and 2 are the
               exponential functions for Al and $^3\mathrm{He}$,
               whereas 3 is the accidental background coincidence.}
      \label{fig:timedis}
\end{figure}
Figure~\ref{fig:timedis} displays the time distributions of Si($dE-E$)
events for the experiment with pure $^{3}\mathrm{He}$.
The time distributions are very well fitted by Eq.~(\ref{eq21}), using
the values $\lambda_{\mathrm{Al}} = 1.156 \times 10^6 \, \mbox
{s}^{-1}$ and $\lambda_{\mathrm {He}} = 0.457\times 10^6 \,
\mbox{s}^{-1}$ from Refs.~\cite{suzuk87,maevx96,acker98}.

The second step is to remove the background arising from muon stops in
aluminum. 
For this purpose, the time interval $\delta t = t_2 - t_1$ is divided
into two subintervals $\delta t_{A}=t_{3} - t_{1}$ and $\delta t_{B}$
= $t_{2} - t_{3}$.
Therefore the proton yields $N_p$ which correspond to the two new
intervals $\delta t_{A}$ and $\delta t_{B}$ have the form
\begin{eqnarray}
      \label{eq22}
      N_{p}^A & = & \int\limits_{t_1 }^{t_3 } \frac{dN_{p} }{dt}dt
      \nonumber \\
      & = & \frac{D_1}{\lambda _{\mathrm {Al}} } e^{ - \lambda
      _{\mathrm {Al}} \cdot t_1} (1 - e^{ - \lambda _{\mathrm {Al}}
      \cdot \delta t_{A}}) \\
      & + & \frac{D_2 }{\lambda_{\mathrm {He}} } e^{ -
      \lambda_{\mathrm {He}} \cdot t_1 } (1 - e^{ - \lambda_{\mathrm
      {He}} \cdot \delta t_{A}}) + C \cdot \delta t_{A}\nonumber
\end{eqnarray}
and 
\begin{eqnarray}
      \label{eq23}
      N_{p}^B & = & \int\limits_{t_3 }^{t_2 } \frac{dN_{p} }{dt}dt
      \nonumber \\
      & = & \frac{D_1 }{\lambda_{\mathrm{Al}} }e^{ - \lambda _{\mathrm
      {Al}} \cdot t_3 }(1 - e^{ - \lambda _{\mathrm {Al}} \cdot
      \delta t_{B} }) \\
      & + & \frac{D_2 }{\lambda_{\mathrm {He}} }e^{ - \lambda_{\mathrm
      {He}} \cdot t_3}(1 - e^{ - \lambda_{\mathrm {He}} \cdot \delta
      t_{B} }) + C \cdot \delta t_B \, .\nonumber
\end{eqnarray}
The total numbers of events $N_{p}^A $ and $N_{p}^B$, given for the
two time intervals $\delta t_{A}$ and $\delta t_{B}$ are given by the
two--dimensional amplitude distributions $(A_{jk})_A$ and
$(A_{jk})_B$, created for each ($jk$) cell, where $j = 1,\ldots, \ell$
and $k = 1,\ldots, m$ are the cell indexes on the $dE$ (the
energy losses in the thin Si detector) and the $E$ axes (the deposited
energy in the thick Si detector), respectively.

The time intervals, $\delta t_{A}$ and $\delta t_{B}$, are chosen
such that the difference between the proton or deuteron yields
measured in the intervals $\delta t_{A}$ and $\delta t_{B}$ is
independent of the aluminum muon capture contribution.
This means that the first parts of Eqs.~(\ref{eq22}) and~(\ref{eq23})
are then equal, i.e.,
\begin{equation}
      \label{eq24}
      \frac{D_1 }{\lambda _{\mathrm {Al}} }e^{ - \lambda _{\mathrm Al}
      \cdot t_3 }(1 - e^{ - \lambda_{\mathrm{Al}} \cdot \delta t_{B}
      }) = \frac{ D_1 }{\lambda _{\mathrm {Al}} }e^{ - \lambda
      _{\mathrm {Al}} \cdot t_1 }(1 - e^{ - \lambda _{\mathrm {Al}}
      \cdot \delta t_{A}}) \, .
\end{equation}

If the initial $t_{1}$ and final $t_{2}$ measurement times are given,
the middle time $t_{3}$ becomes
\begin{equation}
      \label{eq25}
      t_3 = \frac{e^{ - \lambda _{\mathrm
      {Al}} \cdot t_1 } + e^{ - \lambda _{\mathrm {Al}} \cdot t_2
      }}{\lambda _{\mathrm {Al}} }\cdot \ln{2} \, .
\end{equation}

The difference between $N_{p}^B $ and $N_{p}^A $ is the total number
of events in the resulting $(A_{jk})_{B - A}$ two--dimensional
($dE-E$) protons distribution.
This distribution was obtained by subtracting channel by channel the
content of the ($jk$) cell for the two $(A_{jk})_A$ and $(A_{jk})_B$
distributions.

The final number of protons $N_{p}^{\mathrm{final}}$ for the pure
${}^{3}\mathrm{He}$ measurement is then
\begin{eqnarray}
      \label{eq26}
      N_{p}^{\mathrm{final}} & = & N_{p}^B - N_{p}^A
      \nonumber \\
      & = & \frac{N_{\mu}^{\mathrm{He}} \cdot \tilde
      {\varepsilon }_{p} \cdot \lambda _\mathrm{cap}^{p} (\Delta E_{p}
      ) \cdot F_t }{\lambda_{\mathrm{He}} } \\
      & + & C \cdot [\delta t_{B} - \delta t_{A}] \nonumber
\end{eqnarray}
with
\begin{eqnarray}
      \label{eq26a}
      F_t & = & e^{ - \lambda_{\mathrm {He}} \cdot t_1 }(1 - e^{ -
      \lambda_{\mathrm {He}} \cdot \delta t_{A}}) \nonumber \\ & + &
      e^{ - \lambda_{\mathrm {He}} \cdot t_3 }(1 - e^{ -
      \lambda_{\mathrm {He}} \cdot \delta t_{B} }) \, ,
\end{eqnarray}
whereas, for the $\mathrm{D}_2 + {}^{3}\mathrm{He}$ mixture, it
becomes
\begin{eqnarray}
      \label{eq27}
      N_{p}^{\mathrm{final}} & = & N_{\mu}^{\mathrm{D/He}} \cdot
      \tilde {\varepsilon }_{p} \cdot \lambda _\mathrm{cap}^{p}
      (\Delta E_{p} ) \cdot F_t \nonumber \\
      & + & C \cdot [\delta t_{B} - \delta t_{A}] 
\end{eqnarray}
with
\begin{eqnarray}
      \label{eq27a}
      F_t & = & (\mathrm{a}_{\mathrm{He}} + \xi_{\mathrm{D}}) \cdot
       \left.  \frac{2e^{ - \lambda_{\mathrm {He}} \cdot t_3
      } - e^{ - \lambda_{\mathrm {He}} \cdot t_1 } - e^{ -
      \lambda_{\mathrm {He}} \cdot t_2 }}{\lambda_{\mathrm {He}} }
      \right. \nonumber \\
      & - & \xi_{\mathrm{D}}\cdot \left[ \frac{e^{ - \lambda_{\mu
      \mathrm{d}} \cdot t_3 }(1 - e^{ - \lambda_{\mu \mathrm{d}} \cdot
      \delta t_{B} }) }{\lambda_{\mu \mathrm{d}}}  \right. \\
      & - & \left.  \frac{e^{ - \lambda_{\mu \mathrm{d}} \cdot t_1 }(1
      - e^{ - \lambda_{\mu \mathrm{d}} \cdot \delta t_A
      })}{\lambda_{\mu \mathrm{d}} }\right] \, . \nonumber
\end{eqnarray}

Analyzing the data according to Eqs.~(\ref{eq24}) and~(\ref{eq25}) we
obtained the intervals $\Delta t_{A} = [t_1;t_3] =[0.51;1.098]\,
\mu\mbox{s}$ and $\Delta t_{B} = [t_3;t_2]=[1.098;6.0]\,\mu\mbox{s}$.
The corresponding capture events in aluminum amount to $\sim 23$\% of
the total events.
As an example, Table~\ref{tab:background} show the number of events
measured in Run~II in both time intervals and both elements.

\begin{table}[b]
\begin{ruledtabular} 
      \caption{Number of aluminum and helium events in Run~II, as a
               function of the time intervals $\Delta t_{A}$ and
               $\Delta t_{B}$.}
\label{tab:background}
\begin{tabular}{ccccc}
Particle & \multicolumn{2}{c}{Interval  $\Delta t_{A}$}  
& \multicolumn{2}{c}{Interval  $\Delta t_{B}$}  \\ 
& Aluminum & Helium & Aluminum & Helium \\ \hline
Proton & 2700 & 3600 & 2700 & 10\,800 \\
Deuteron & 1150 & 1650 & 1150 & 5800 \\
\end{tabular}
\end{ruledtabular}
\end{table}

Our subtraction method, while reducing the number of events in helium
by a factor 2 (see Table~\ref{tab:background}), yields essentially
background--free events.
However, Eqs.~(\ref{eq26}) and~~(\ref{eq27}) still contain some parameters
that need to be determined, namely the energy interval $\Delta E_{p}$
and the accidental coincidence background described by the constant
$C$.

\begin{figure}[ht]
      \includegraphics[width=7.8cm,angle=90]{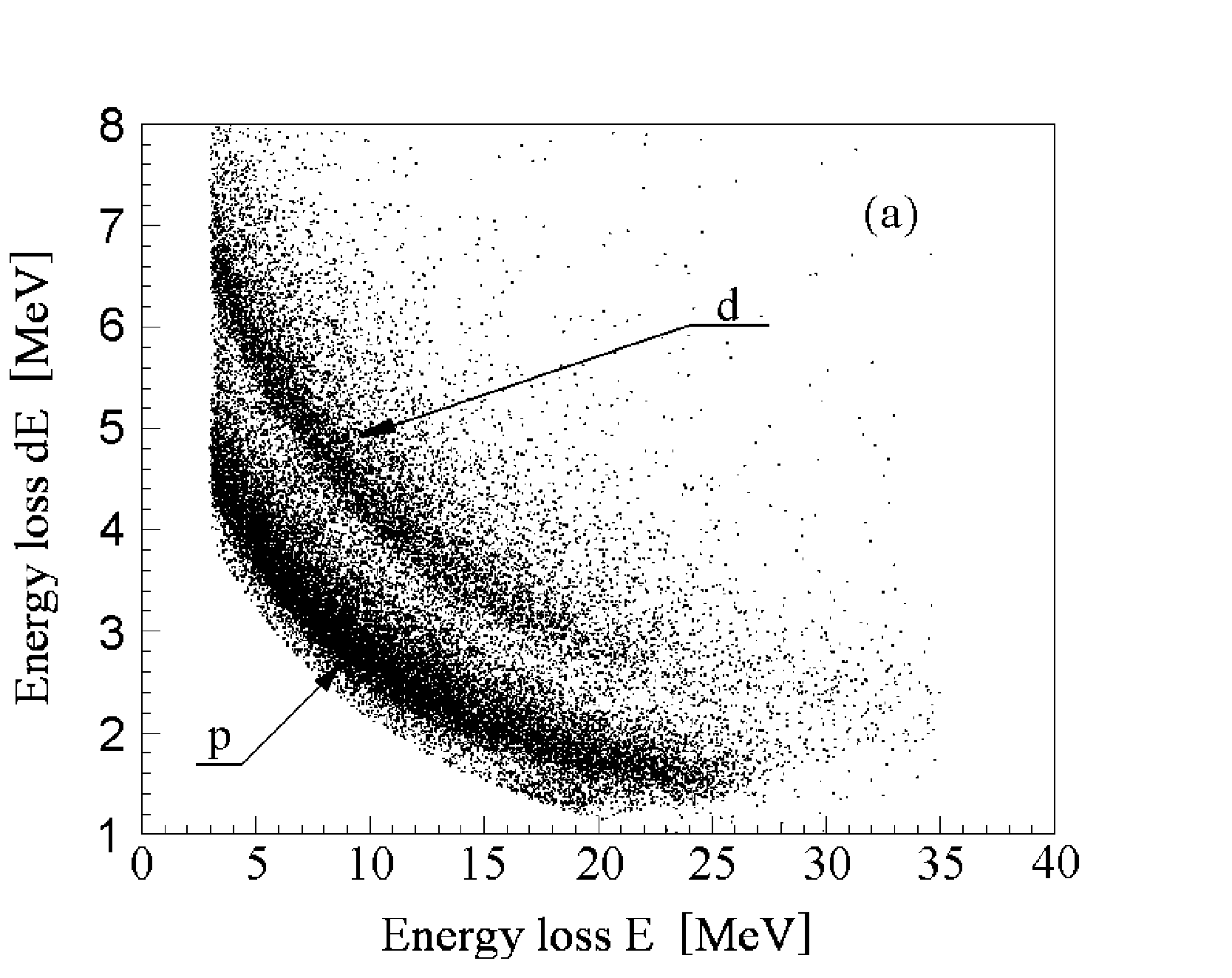}
      \includegraphics[width=7.8cm,angle=90]{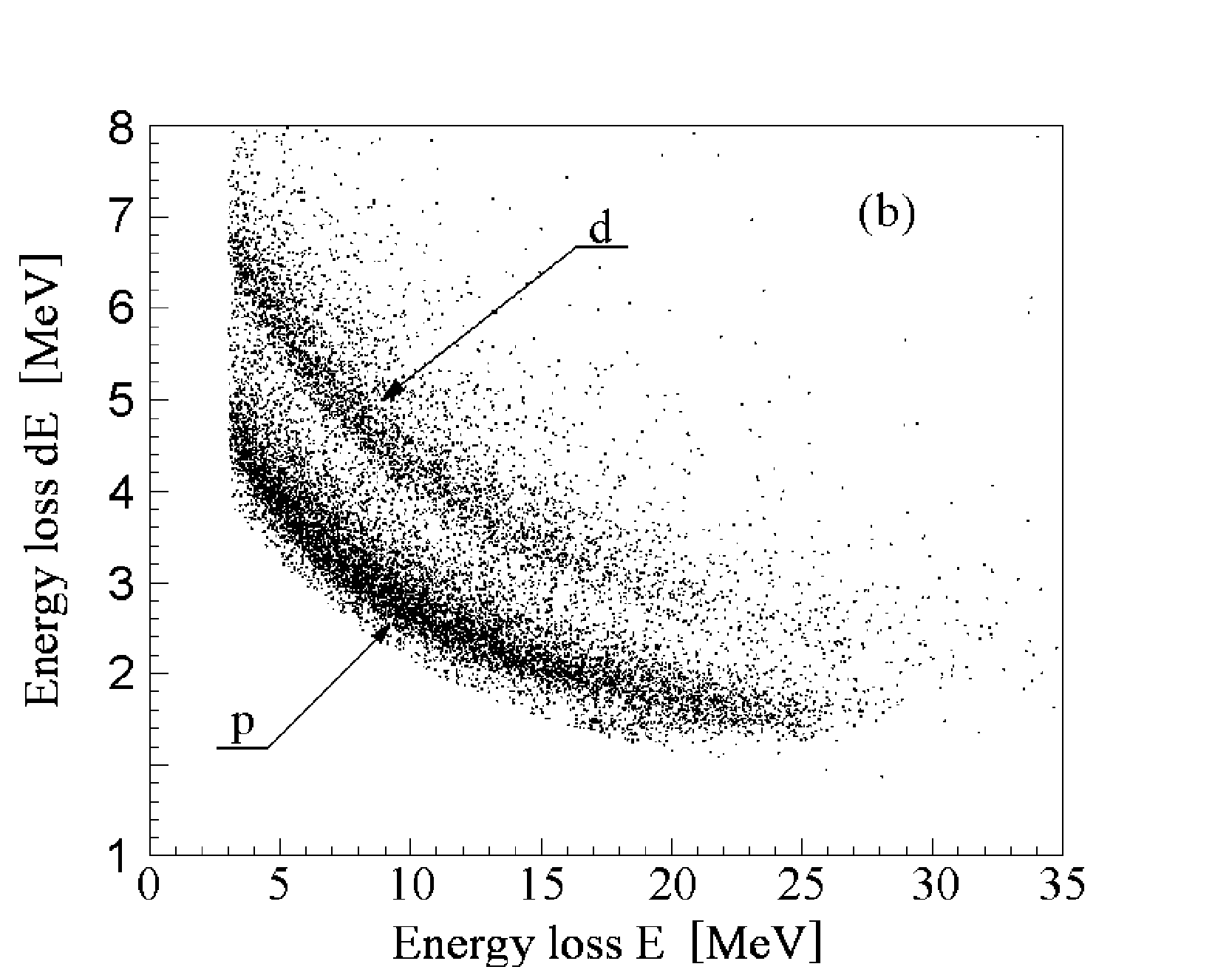}
      \caption{Two--dimensional energy distributions of the Si($dE-E$)
               detector events for the time interval $0.5 \leq t \leq
               6.0$~$\mu$s relative to the muon stop.  (a) represents
               Run~I (pure $^3\mathrm{He}$) and (b) Run~II
               ($\mathrm{D}_2+{}^3\mathrm{He}$ mixture).  The
               distinction between protons and deuterons is clearly
               visible for both measurements.}
\label{fig:scat}
\end{figure}

The energy intervals for detecting protons and deuterons by the
Si($dE-E$) detectors were chosen such that the real detection
sensitivity is the same for any initial energies.
This allows us to remove any possible distortion in our amplitude
distribution, which would occur for too low or too high energies.
The chosen limits are $4 - 23$~MeV for both protons and deuterons in
the thick $E$ detector.
The thin $dE$ detector has two different energy intervals, namely $1 -
6$~MeV for the protons and $2 - 8$~MeV for the deuterons.

Figure \ref{fig:scat} displays the two--dimensional ($dE-E$)
distributions of events detected by the Si($dE-E$) detectors in Run~I
with pure ${}^{3}\mathrm{He}$ and in Run~II with the $\mathrm{D}_2 +
{}^{3}\mathrm{He}$ mixture.
The two distinct branches of events corresponding to the protons
and the deuterons are clearly visible and lie inside our chosen energy
intervals.
Note that the shapes of the two--dimensional
($dE-E$) distributions obtained in the runs with pure
${}^{3}\mathrm{He}$ and with the $\mathrm{D}_2 + {}^{3}\mathrm{He}$
mixture coincide.
This indicates that there are no neglected systematic errors, and that
the algorithm used for the data analysis is correct.

As to the accidental coincidence background described by the constant
$C$, its contribution to Eqs.~(\ref{eq26}) and~(\ref{eq27}) is small
when compared to the muon stop contributions in Al and
${}^{3}\mathrm{He}$, as can be seen in Fig.~\ref{fig:timedis}.
The constant $C$ was quantitatively determined in each Run by fitting
the time distribution, as given in Fig.~\ref{fig:timedis}, including the
time interval $-0.4 \, \mu \mbox{s} \le t \le 0$ with respect to the
muon stop.
Details of such a fit shown in the muon decay electron time spectra
are in Fig.~\ref{fig:electron}.

%Runs~(I--III) from the
%two--dimensional Si($dE-E$) distributions analysis in the 
%time distribution 

As mentioned in the introduction, we want to determine different
characteristics of the muon capture by ${}^{3}\mathrm{He}$ nuclei, namely:
\begin{itemize}
  \item[--] the initial energy distributions of protons and deuterons
($S(E_p),\ S(E_d)$),
  \item[--] the muon capture rates as function of the energy for both
the protons and deuterons ($\lambda _\mathrm{cap}^p (\Delta E_p )$ and
$\lambda _\mathrm{cap}^d (\Delta E_d )$),
  \item[--] their derivatives $d\lambda _\mathrm{cap}^p / dE_p$ and
$d\lambda _\mathrm{cap}^d / dE_d$.
\end{itemize}
For this purpose, following Eqs.~(\ref{eq9}), (\ref{eq20}), and
(\ref{eq8}), we need to determine the number of protons $N_{p}(\Delta
E_{p}$, $\Delta T)$ and deuterons $N_{d}(\Delta E_{d}$,
$\Delta T)$ for each energy interval $\Delta E_{p}$ and
$\Delta E_{d}$.
In the next two subsections we describe the two approaches to
determine the respective number of protons and deuterons, based on the
analysis of the two--dimensional $(A_{jk})_{B - A}$ distributions as
function of $dE$ and $E$ for each of the three Runs~(I--III).

\subsection{Method I: Least--squares}
\label{sec:method-i:-least}

The principle of this method is to use Monte Carlo (MC) simulations to
reproduce the experimental data and to minimize the free parameters
which are required by such a simulation.
The simulation conditions and parameters will be given below.
The energy spectra of the protons and deuterons produced by
reactions~(\ref{eq1}) and~(\ref{eq2}) are divided into $i$
subintervals of 1~MeV fixed widths.
Since the theoretical maximum energies are $\approx 53$~MeV for the
protons and $\approx 33$~MeV for the deuterons, the numbers of
subintervals are 53 and 33, respectively.

Using the experimental muon stop distribution in our target, we
simulate the probability $P^{\mathrm {MC}}(A_{jk} / E_{p}^i )$ that a
proton (analogously a deuteron) produced with an energy $E_{p}^i$ (in
the $i$-th interval $\Delta E_p^i$) will be detected by the Si($dE-E$)
detectors in the ($jk$) cell of the two--dimensional distribution
$A_{jk}$.
This probability is
\begin{equation}
      \label{eq28}
      P^{\mathrm {MC}}(A_{jk} / E_{p}^i ) = \frac{(n_{jk} )_i^{\mathrm
      {MC}} }{(n_o)_i },
\end{equation}
where $(n_{jk})_i^{\mathrm {MC}}$ is the number of simulated events
detected in the ($jk$) cell when the number of protons, which were
created with an initial energy $E_{p}^i$ from the interval $\Delta E_p^i$, 
is $(N_o)_i$.
The ($jk$) cell size is chosen arbitrarily and mainly depends on the
statistics of events $(n_{jk})_i^{\mathrm {MC}}$ belonging to a particular 
cell ($jk$).

Then the MC simulated ``pseudo--experimental'' (i.e., normalized to the
experimental counts $N_p^{\mathrm{final}} = N_{p}(\Delta E_{p},\Delta
T)$) event numbers $(N_{jk})^{\mathrm {MC}}$ for each ($jk$) cell,
become
\begin{eqnarray}
      \label{eq28a}
      (N_{jk} )^{\mathrm {MC}} & = & N_{p}^{\mathrm{final}} 
      \sum\limits_i P^{MC}(A_{jk}/E^i_{p}) \nonumber \\
      & \times & \int\limits _{E^i_p}^{E^i_p+\delta E^i_p} 
      S(E_p)dE_p \, , 
\end{eqnarray}
where $S(E_p)$ is the initial proton energy distribution normalized to unity
in the full energy interval $\Delta E_p$. 

In our energy intervals $10 \le E_p \le 49$~MeV and $13 \le E_{d} \le
31$~MeV both the proton and deuteron energy distributions, $S(E_{p})$
and $S(E_{d})$,
obtained via the impulse approximation model and the realistic wave
functions for the ${}^{3}\mathrm{He}$ nucleus ground
state~\cite{phili75,glock90,skibi99,glock96} can be well described by
the expression
\begin{equation}
      \label{eq29}
      S(E_{p}) = A_{p} e^{ - \alpha _{p} E_{p} },
\end{equation}
where the amplitude $A_{p}$ and the fall--off yield $\alpha_{p}$ 
are the variable parameters. Thus Eq.~(\ref{eq28a}) can be 
rewritten as
\begin{equation}
      \label{eq30}
      (N{jk})^{MC} = N_{p}^\mathrm{final} \sum\limits_i 
      P^{MC}(A_{jk}/E^i_{p}) \cdot \tilde{S}(E_p^i) \, , 
\end{equation}
where 
\begin{equation}
      \label{eq30a}
      \tilde{S}(E_p^i) = 
      \frac {1 - e^{-\alpha _{p}\cdot \delta E_p^i}}
      {\alpha _{p}} S(E_p^i)\, . 
\end{equation}

We created $(N_{jk})^{MC} $ for different values of the amplitude
$A_{p}$ and the fall--off yield $\alpha_{p}$ and used the $\chi ^{2}$
minimization procedure between the MC and experimental events,
\begin{equation}
      \label{eq33}
      \chi ^2 = \sum\limits_{j =  1}^l {\sum\limits_{k = 1}^m
      {\frac{\left[ ({N_{jk})^{\exp } - (N_{jk} )^{\mathrm {MC}}}
      \right]^2}{\sigma _{N_{jk}^{\exp } }^2 }} } ,
\end{equation}
to obtain the best values for the parameters $A_{p}$ and
$\alpha_{p}$ which describe the initial energy distribution of protons
$S(E_p)$.  $(N_{jk})^{\exp}$ is the number of
measured events belonging to the ($jk$) cell, obtained for
%by Eqs.~(\ref{eq26}) and~(\ref{eq27}) for
pure $^{3}\mathrm{He}$ and the $\mathrm{D}_2 + {}^{3}\mathrm{He}$
mixture, respectively.
The $N_{p}^{\mathrm{final}}$ values of  Eqs.~(\ref{eq26}) and
(\ref{eq27}) represent the sum of the $(N_{jk})^{\exp}$ over $j$ and
$k$.

\begin{figure}[t]
      \includegraphics[angle=90,width=0.46\textwidth]{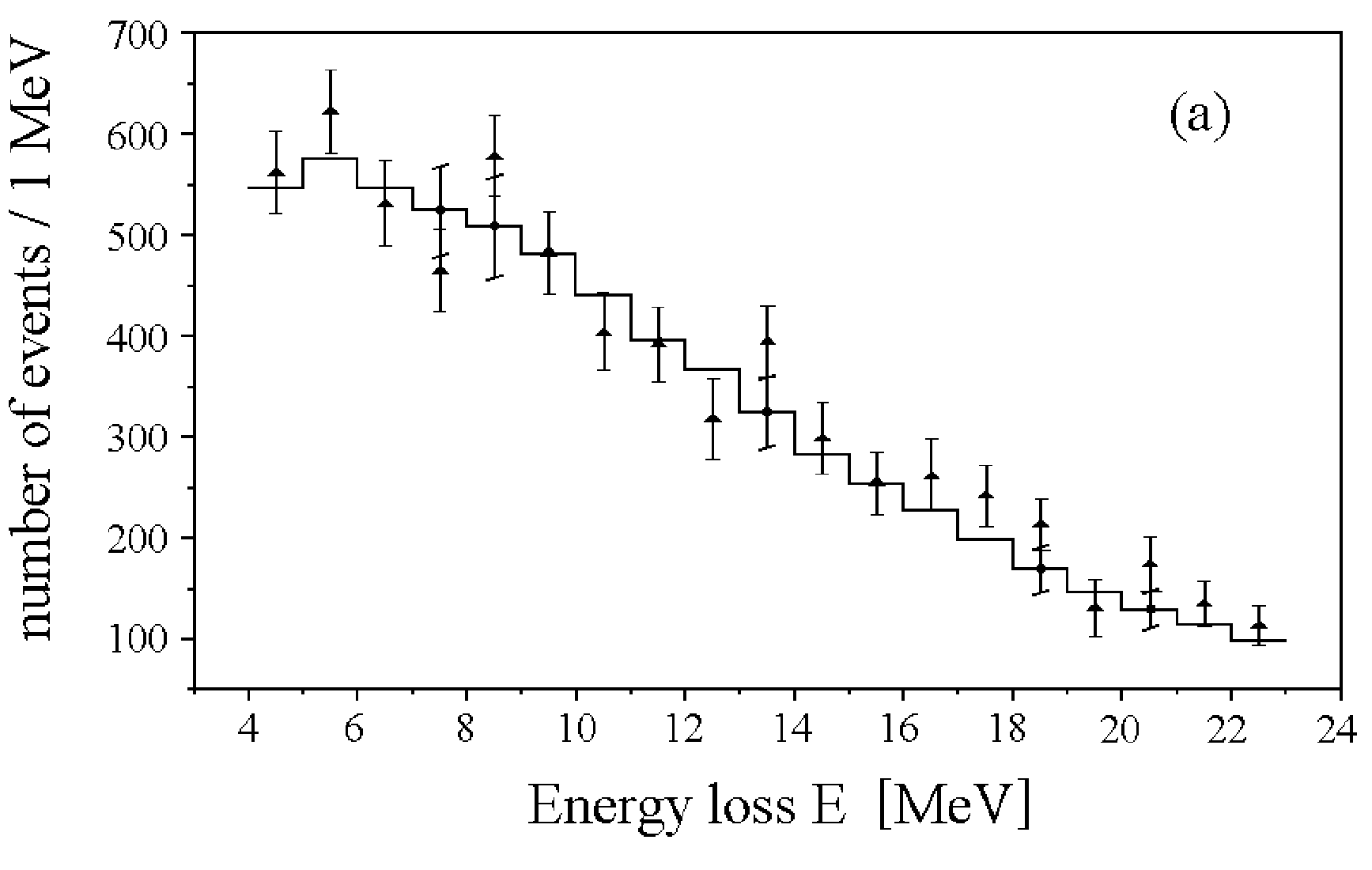}
      \includegraphics[angle=90,width=0.46\textwidth]{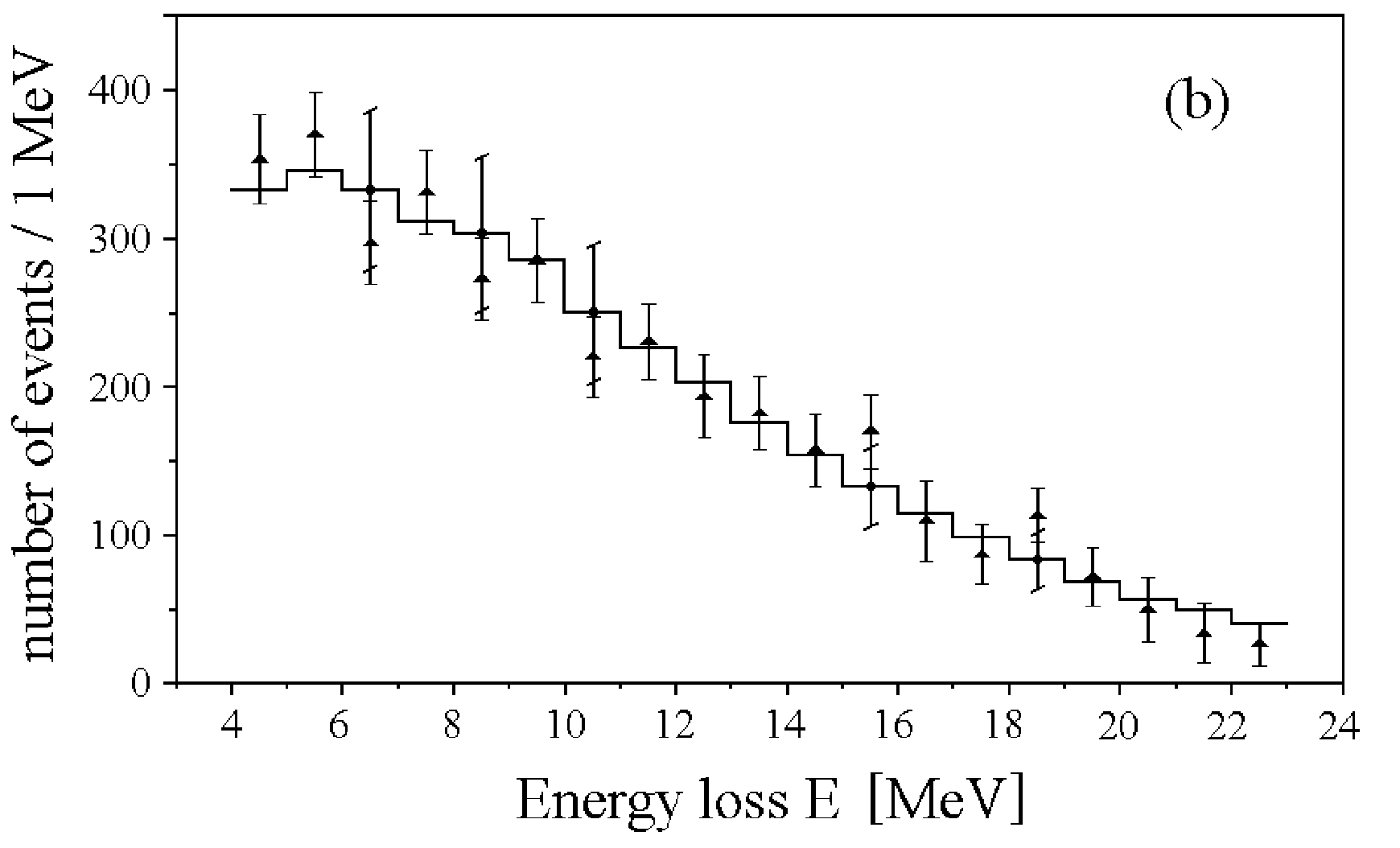}
      \caption{$E$ axis projections of the two--dimensional ($dE-E$)
               distributions for the protons (a) and the deuterons (b)
               obtained in Run~II with the $\mathrm{D}_2 +
               {}^{3}\mathrm{He}$ mixture.  The histogram shows the
               experimental data, whereas the black triangles are MC
               events from Method I\@.}
      \label{fig:distr1}
\end{figure}

A second and parallel minimization is done when projecting the
experimental and MC events onto the two energy axes $j$ and $k$.
When projecting onto the $E$ axis, we have the experimental data as
\begin{equation}
      \label{eq34}
      (N_k)^{\exp } = \sum\limits_{j = 1}^l {(N_{jk})^{\exp } } ,
\end{equation}
and the MC events as
\begin{eqnarray}
      \label{eq35}
      (N_k )^{\mathrm {MC}} &= &\sum\limits_{j = 1}^l {(N_{jk}
      )^{\mathrm {MC}}} \\ 
      & = & N_{p}^\mathrm{final} \sum\limits_i {\sum\limits_{j = 1}^l 
      {P^{MC}(A_{jk} / } } E_{p}^i ) \cdot \tilde{S}(E^i_p) \, .
      \nonumber
\end{eqnarray}
Therefore $\chi^2$ becomes
\begin{equation}
      \label{eq36}
      \chi ^2 = \sum\limits_{k = 1}^m {\frac{\left[ {(N_k)^{\exp } -
      (N_k )^{\mathrm {MC}}} \right]^2}{\sigma _{N_k^{\exp } }^2 }} \,
      .
\end{equation}
Similar equations can be written for the second axis $j$ when we
project the events onto the $dE$ axis.

Figure~\ref{fig:distr1} displays the least--squares comparison of the
$E$ axis projection of the two--dimensional experimental and the MC
simulated distributions for the protons and the deuterons of Run~II.
As seen, the MC distributions correspond very well to the experimental
proton and deuteron energy distributions, thus strongly supporting our
analysis method I\@.

The amplitude and fall--off yield results from the three experimental
Runs~(I--III) are
\begin{eqnarray}
      \label{eq36a}
      {A}_p =& (0.832 & \pm \,\, 0.043 )\,\, \mbox{MeV}^{ - 1} \nonumber
      \\
      \alpha_p =& (-0.163 & \pm \,\, 0.002) \,\, \mbox{MeV}^{ - 1}
\end{eqnarray}
for the protons and 
\begin{eqnarray}
      \label{eq36b}
      {A}_d =& (5.59 &\pm \,\, 1.39) \,\, \mbox{MeV}^{ - 1} \nonumber \\
      \alpha_d =& (-0.243 &\pm \,\, 0.012) \,\, \mbox{MeV}^{ - 1}
\end{eqnarray}
for the deuterons.

\begingroup
      \squeezetable
\begin{table}[t]
\begin{ruledtabular} 
      \caption{Mean proton energy distribution normalized to unity in
               the energy range $10 \le \mathrm{E}_{p} \le 49$~MeV,
               from methods I and II\@.}
\label{tab:p-energy}
\begin{tabular}{ccc}
$E_{p}$ & \multicolumn{2}{c}{$<S(E_p)>$ [MeV$^{-1}$]} \\ 
{}[MeV]& Method I & Method II \\ \hline 
10.5  & 0.150 (13)   & 0.1570 (83)   \\
11.5  & 0.127 (12)   & 0.1309 (48)   \\
12.5  & 0.108 (10)   & 0.1077 (35)   \\
13.5  & 0.0922 (88)  & 0.0958 (29)   \\
14.5  & 0.0784 (77)  & 0.0765 (24)   \\
15.5  & 0.0667 (67)  & 0.0644 (22)   \\
16.5  & 0.0568 (59)  & 0.0525 (22)   \\
17.5  & 0.0483 (51)  & 0.0485 (20)   \\
18.5  & 0.0411 (45)  & 0.0392 (18)   \\
19.5  & 0.0349 (39)  & 0.0313 (17)   \\
20.5  & 0.0297 (34)  & 0.0284 (16)   \\
21.5  & 0.0252 (30)  & 0.0251 (14)   \\
22.5  & 0.0215 (26)  & 0.0208 (14)   \\
23.5  & 0.0183 (22)  & 0.0184 (13)   \\
24.5  & 0.0155 (20)  & 0.0162 (11)   \\
25.5  & 0.0132 (17)  & 0.0150 (12)   \\
26.5  & 0.0112 (15)  & 0.01135 (28)  \\
27.5  & 0.0095 (13)  & 0.00934 (19)  \\
28.5  & 0.0081 (11)  & 0.00793 (16)  \\
29.5  & 0.00688 (98) & 0.00679 (14)  \\
30.5  & 0.00585 (85) & 0.00585 (22)  \\
31.5  & 0.00497 (74) & 0.00528 (27)  \\
32.5  & 0.00422 (64) & 0.00440 (21)  \\
33.5  & 0.00359 (56) & 0.00379 (16)  \\
34.5  & 0.00305 (49) & 0.00316 (13)  \\
35.5  & 0.00259 (42) & 0.00255 (11)  \\
36.5  & 0.00220 (37) & 0.00210 (9)   \\
37.5  & 0.00187 (32) & 0.00177 (7)   \\
38.5  & 0.00158 (28) & 0.00142 (6)   \\
39.5  & 0.00135 (24) & 0.00119 (5)   \\
40.5  & 0.00114 (21) & 0.00105 (4)   \\
41.5  & 0.00097 (18) & 0.00092 (4)   \\
42.5  & 0.00082 (16) & 0.00079 (3)   \\
43.5  & 0.00070 (13) & 0.00067 (3)   \\
44.5  & 0.00059 (12) & 0.00057 (2)   \\
45.5  & 0.00050 (10) & 0.00048 (2)   \\
46.5  & 0.00043 (9)  & 0.00041 (2)   \\
47.5  & 0.00036 (8)  & 0.00034 (1)   \\
48.5  & 0.00031 (7)  & 0.00029 (1)   \\
\end{tabular}
\end{ruledtabular}
\end{table}
\endgroup

The capture rates $\lambda _\mathrm{cap}^{p} (\Delta E_{p} )$ are obtained
after using Eq.~(\ref{eq30}) to calculate the proton yield
$N_{p}(\Delta E_{p}, \Delta T)$ and then applying Eqs.~(\ref{eq9})
and~(\ref{eq20}).
The differential capture rates $d\lambda^p_\mathrm{cap} / dE_p$ 
also follow from the proton yield and  Eq.~(\ref{eq8}); they
are given in Figs.~\ref{fig:capt1} and~\ref{fig:capt2} for the
protons and deuterons, respectively.

\begin{figure}[hb]
      \includegraphics[angle=90,width=0.47\textwidth]{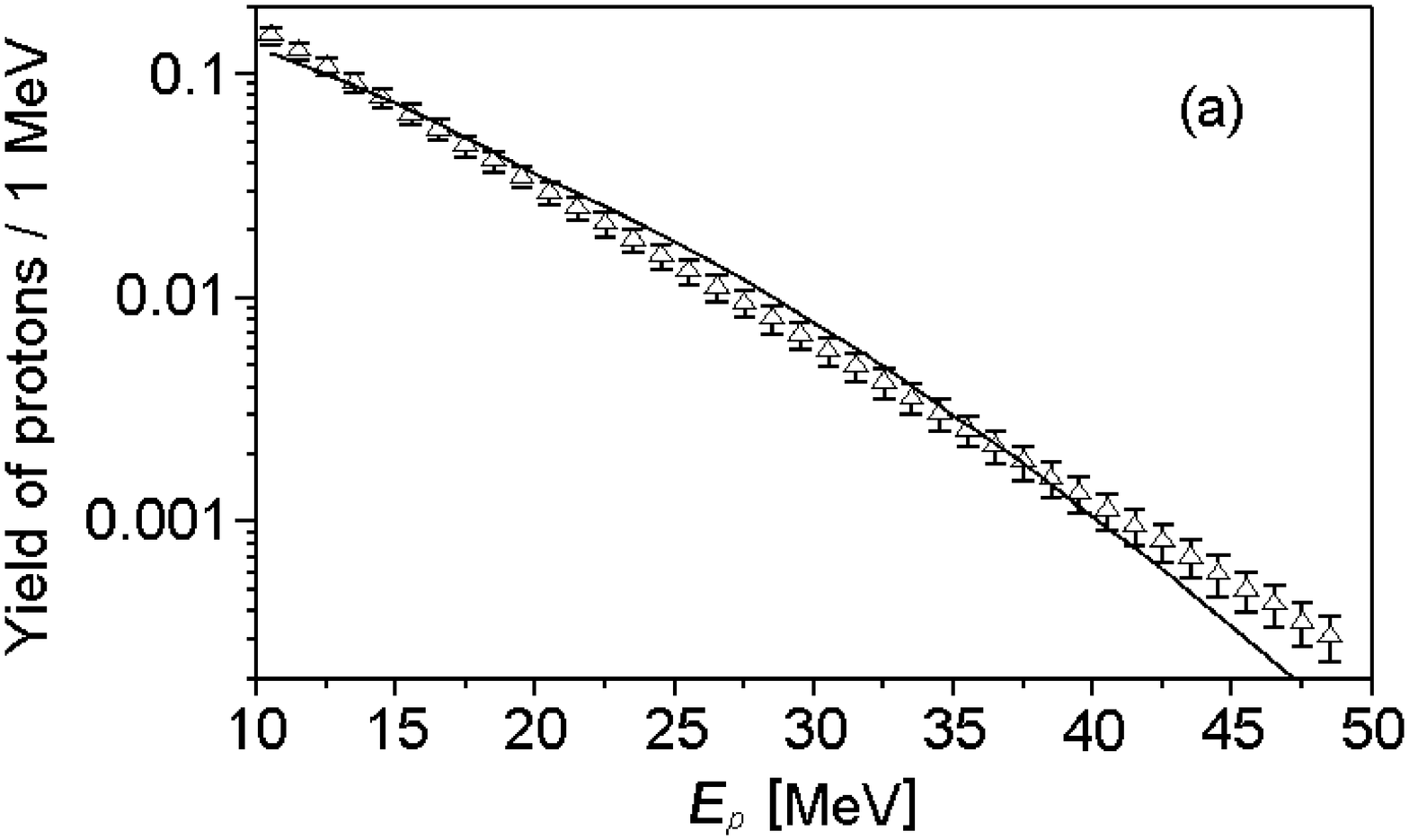}
      \includegraphics[angle=90,width=0.47\textwidth]{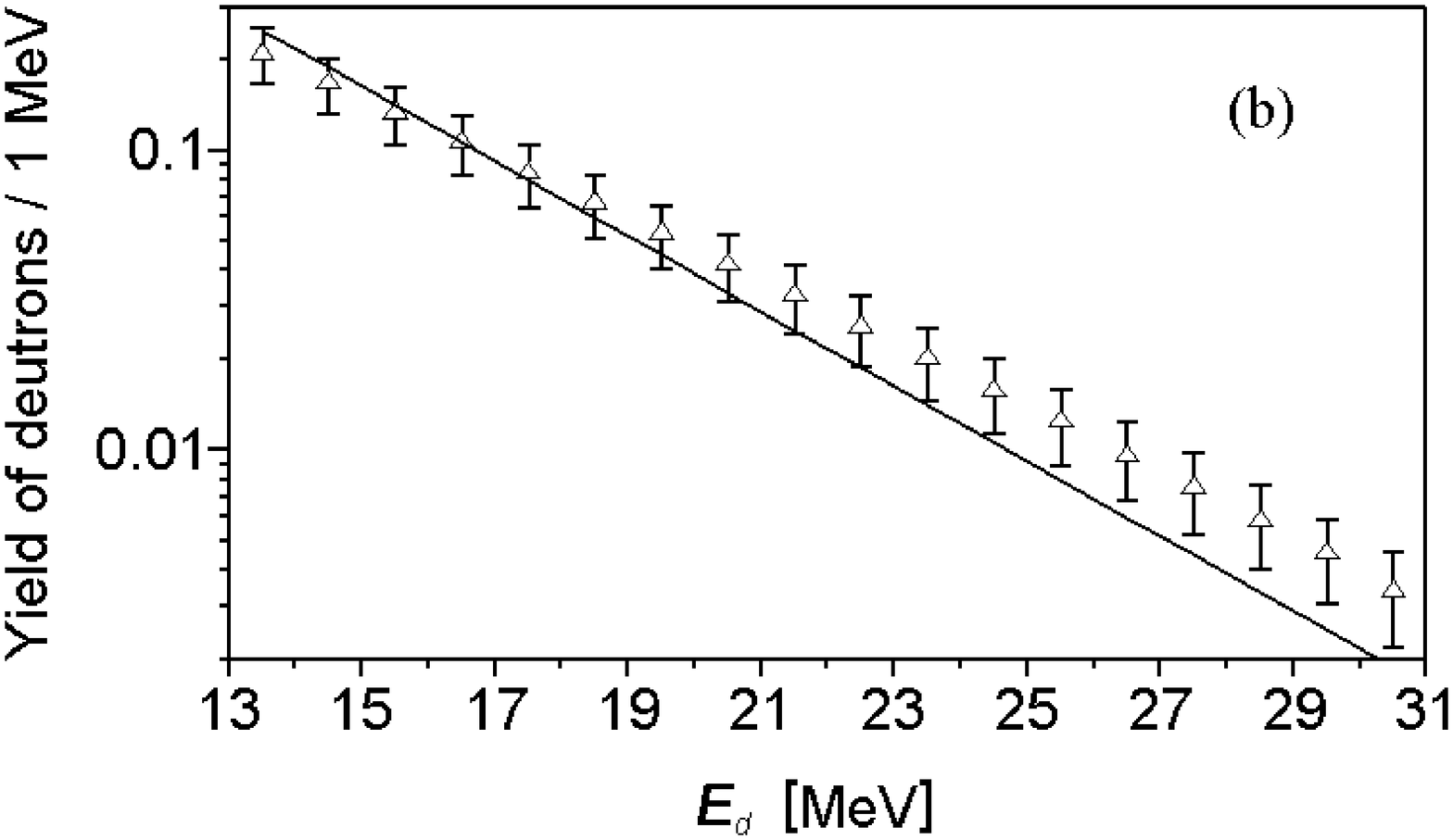}
      \caption{Experimental energy distributions (open triangles) of
               protons (a) and deuterons (b) found by the method I and
               averaged over Runs~(I--III) in comparison with the
               theoretical model~\cite{phili75} (solid line).}
      \label{fig:energy}
\end{figure}

The average energy distributions $<S(E_p>$ and $<S(E_d)>$ from
Runs~(I--III) normalized to unity for the energy intervals $ 10 \le
E_{p} \le 49$~MeV and $13 \le E_{d} \le 31$~MeV are given in
Table~\ref{tab:p-energy} for the protons and in
Table~\ref{tab:d-energy} for the deuterons.
Figure~\ref{fig:energy} displays the energy distributions $<S(E_p)>$
and $<S(E_d)>$ averaged over Runs~(I--III) in comparison with the
model distributions obtained when treating the muon capture in the
simple plane--wave impulse approximation~\cite{phili75} and in the
impulse approximation with the realistic Bonn B
potential~\cite{machl89} of NN interaction in the final
state~\cite{skibi99}.

\begingroup
  \squeezetable
\begin{table}[t]
\begin{ruledtabular} 
      \caption{Mean deuteron energy distribution normalized to unity
               in the energy range $13 \le E_{d} \le 31$~MeV, from
               methods I and II\@.}
\label{tab:d-energy}
\begin{tabular}{ccc}
$E_{d}$ &\multicolumn{2}{c}{$<S(E_d)>$ [MeV$^{-1}$]} \\ 
{}[MeV]& Method I & Method II \\ \hline 
13.5  & 0.210 (44)  & 0.216 (11)  \\
14.5  & 0.167 (36)  & 0.1690 (65) \\
15.5  & 0.133 (29)  & 0.1281 (47) \\
16.5  & 0.106 (24)  & 0.1043 (41) \\
17.5  & 0.084 (19)  & 0.0842 (35) \\
18.5  & 0.067 (16)  & 0.0674 (29) \\
19.5  & 0.053 (13)  & 0.0521 (25) \\
20.5  & 0.042 (10)  & 0.0426 (23) \\
21.5  & 0.0328 (84) & 0.0345 (23) \\
22.5  & 0.0258 (68) & 0.0251 (21) \\
23.5  & 0.0202 (55) & 0.0181 (18) \\
24.5  & 0.0158 (44) & 0.0132 (18) \\
25.5  & 0.0124 (35) & 0.0124 (17) \\
26.5  & 0.0096 (28) & 0.0101 (16) \\
27.5  & 0.0075 (23) & 0.0071 (16) \\
28.5  & 0.0058 (18) & 0.0058 (19) \\
29.5  & 0.0045 (14) & 0.0052 (18) \\
30.5  & 0.0034 (12) & 0.0044 (14) \\
\end{tabular}
\end{ruledtabular}
\end{table}
\endgroup

Experimental and theoretical results agree quite well within the
statistical errors for the energy ranges $10 \le E_{p} \le 40$~MeV and
$13 \le $ $E_{d} \le 24$~MeV, respectively.
For proton energies $E_{p} > 40$~MeV and deuteron energies $E_{d} >
24$~MeV a discrepancy exceeding the tolerable range determined by the
statistical errors is observed.
The cause of the discrepancy is not clear yet. 
It may be due to the necessity of taking into account exchange current
contributions in the interaction and nucleon pair correlations in muon
capture by the ${}^{3}\mathrm{He}$ nucleus.

\subsection{Method II: Bayes theorem}
\label{sec:method-ii:-bayes}

\begin{figure}[t]
      \includegraphics[angle=90,width=0.45\textwidth]{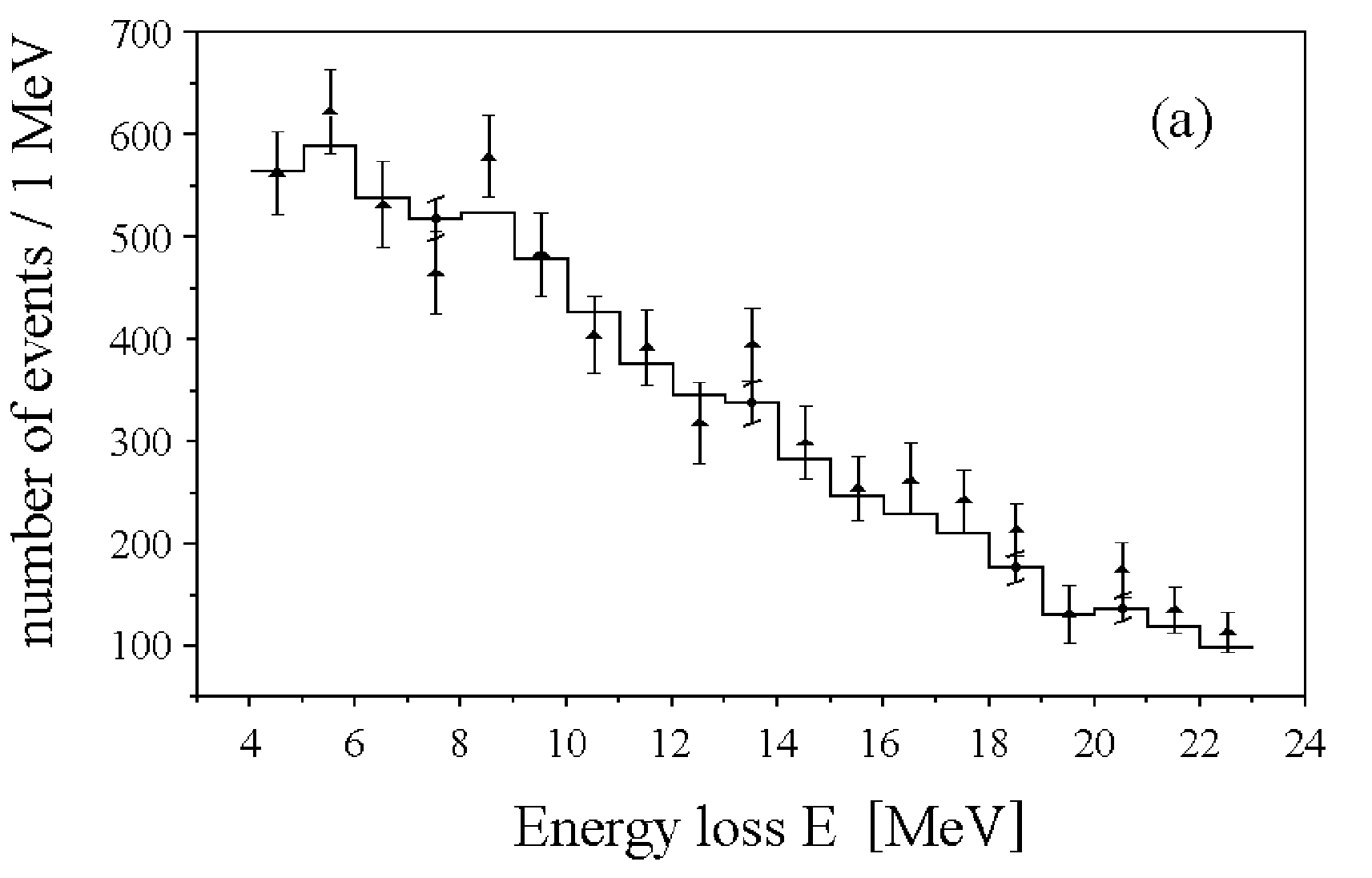}
      \includegraphics[angle=90,width=0.45\textwidth]{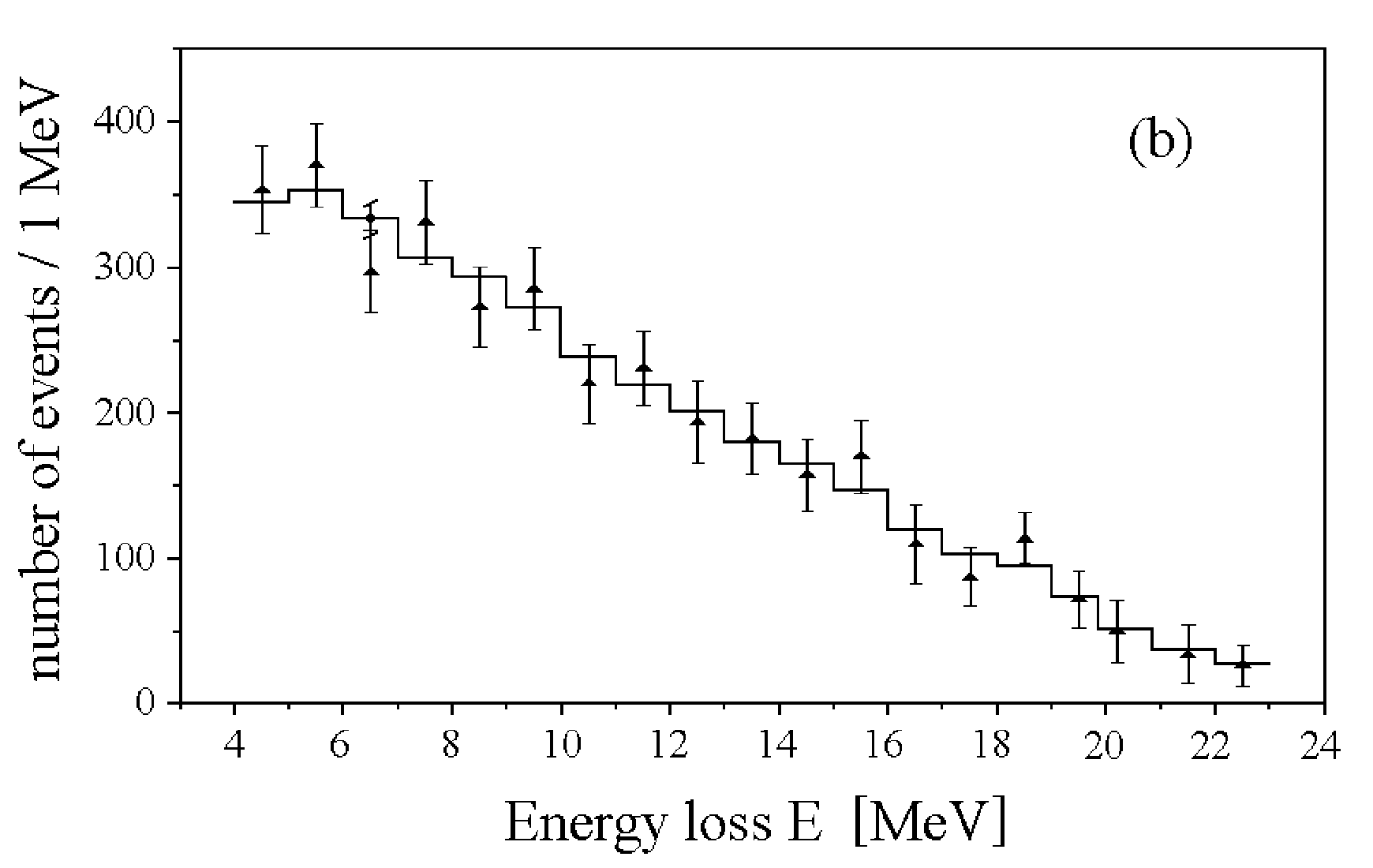}
      \caption{$E$ axis projections of the two-dimensional ($dE-E$)
               distributions for the protons (a) and deuterons (b)
               obtained in Run~II\@.  The histogram shows the
               experimental data, whereas the black triangles are MC
               events from Method II\@.}
      \label{fig:distr2}
\end{figure}

In this approach we use the Bayes
theorem~\cite{gnede61,johns64,boxxx92,agost95,agost99} to determine
the initial energy distribution, $S(E)$, of the protons and the
deuterons produced by muon capture in ${}^{3}\mathrm{He}$.
For this purpose, we apply inverse transformations 
from the  detected two dimensional ($dE - E$) amplitude distributions.

The relation between the probability $P(A_{jk} / E_{p}^i)$ that a proton 
produced with an initial energy $E_{p}^i$ (in the $i$--th interval of 
1~MeV width in our case) will be detected by the Si($dE-E$) telescopes 
and the inverse 
probability $P(E_{p}^i / A_{jk})$ (probability that a proton detected in the 
($jk$) cell comes from the $\Delta E^i_p$ subinterval) is
\begin{equation}
      \label{eq39}  
      P(E_{p}^i / A_{jk}) = \frac{\tilde{S}(E^i_p)\cdot P(A_{jk} / 
      E_{p}^i)} {\sum \limits_i \tilde{S}(E^i_p)\cdot P(A_{jk} / E_{p}^i)}\, .
\end{equation}
The probability $P(A_{jk} / E_{p}^i)$ is given by the MC
simulated probability $P^{MC}(A_{jk} / E_{p}^i)$ defined in
Eq.~(\ref{eq28}).

In the first step of the analysis we start from the initial energy
distribution $S_o(E_p) = S(E_p)$ given by~(\ref{eq29}) with an 
arbitrary set of parameters.
When using the probability given by Eq.~(\ref{eq39})  and the experimental 
data of each ($jk$) cell, we obtain a set of $i$ relations
\begin{equation}
      \label{eq38}
      N_{p}(\Delta E_{p},\Delta T) \cdot \tilde{S}(E_{p}^i ) =
      \frac{\sum\limits_{j = 1}^{l}{\sum\limits_{k = 1}^{m} {P(E_{p}^i /
      A_{jk} )} \cdot (N_{jk})^{\exp } }}{P(A / E_{p}^i) }\, ,
\end{equation}
where $N_{p}(\Delta E_{p}, \Delta T)$ corresponds to Eqs.~(\ref{eq7})
or~(\ref{eq19}), and $P(A/E_{p}^i)$ is the probability that a proton
of initial energy $E_{p}^i$ is detected anywhere in the proton branch
of the two--dimensional distribution $A_{jk}$.
This probability can be written as
\begin{equation}
      \label{eq41}
      P(A / E_{p}^i ) = \sum\limits_{j = 1}^l {\sum\limits_{k = 1}^m
      {P^{MC}(A_{jk} / E_{p}^i )} } \, .
\end{equation} 
We then compare $N_{p}(\Delta E_{p}, \Delta T)$ and the experimental
counts \(N^{\exp} = \sum \sum (N_{jk})^{\exp } \)
for each $i$-th interval via a $\chi^2$ analysis and obtain a proton
energy distribution $S(E_{p}^i )$ from Eqs.~(\ref{eq38}) and 
(\ref{eq30a}).
As long as the $\chi^2$ is not satisfactory, we re--use 
the last $\tilde{S}(E_{p}^i )$ as the starting values in
Eq.~(\ref{eq39}) in the next iteration.

\begin{figure}[t]
      \includegraphics[angle=90,width=0.47\textwidth]{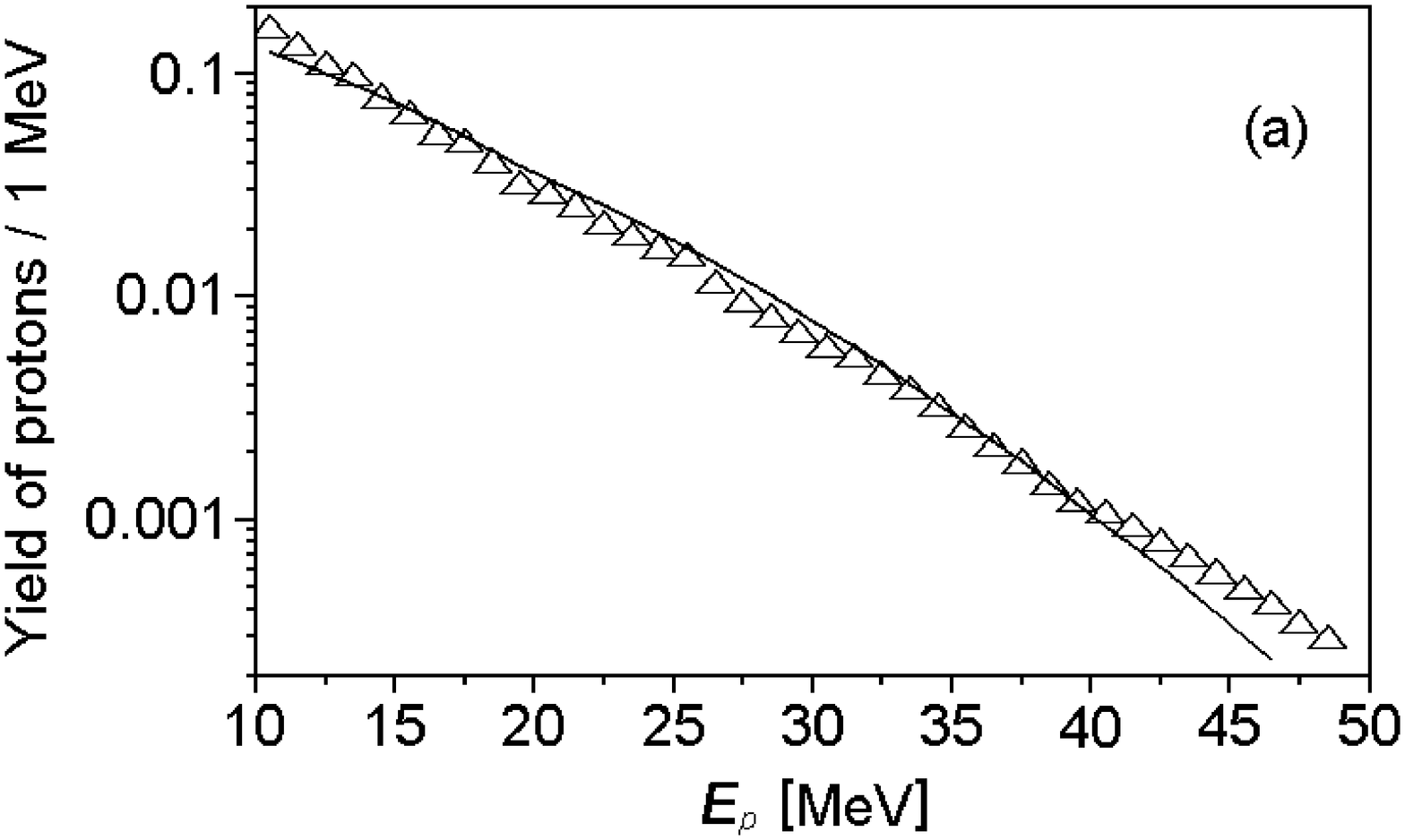}
      \includegraphics[angle=90,width=0.47\textwidth]{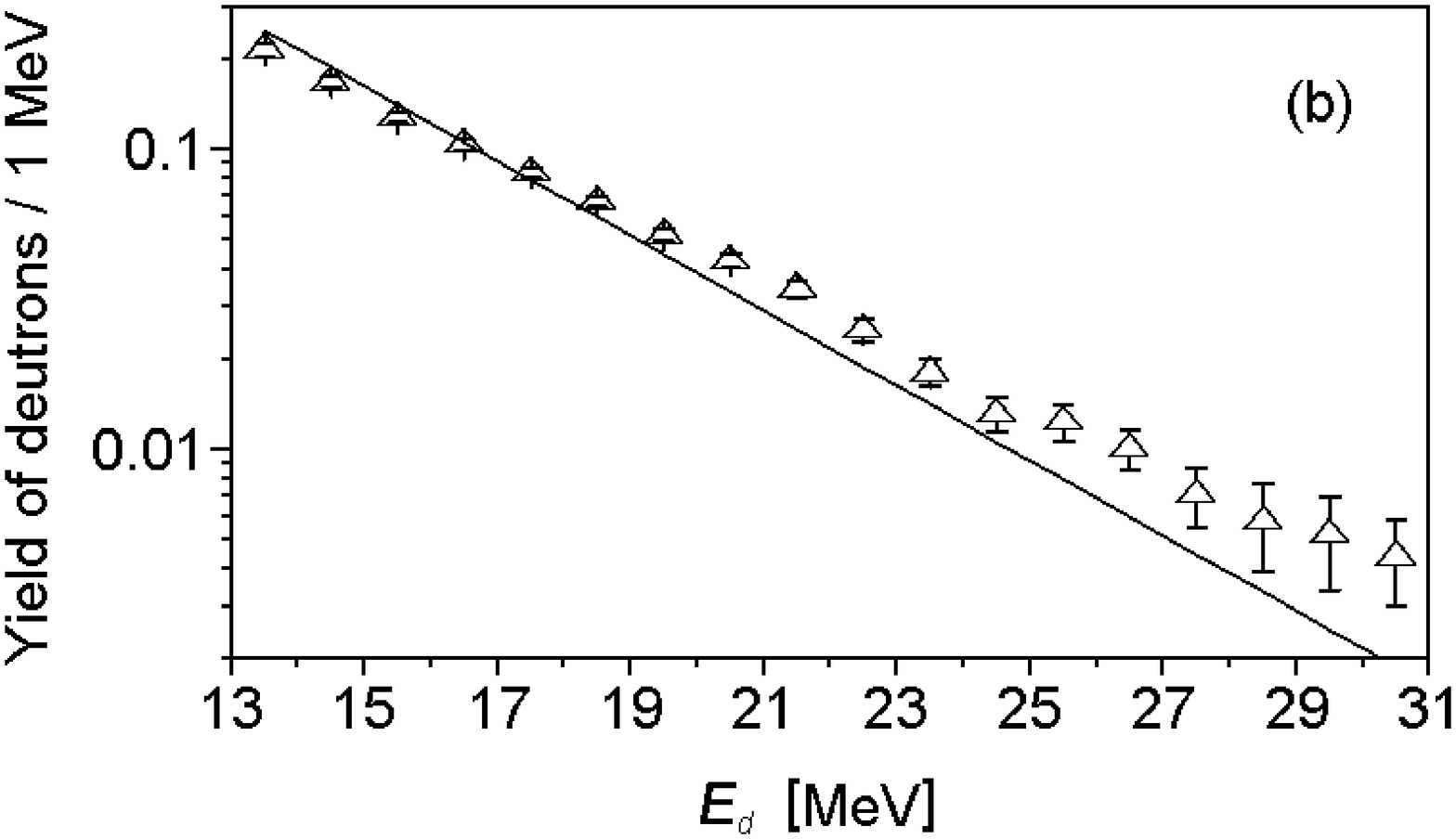}
      \caption{Experimental energy distributions (open triangles) of
               protons (a) and deuterons (b) found by the method II
               and averaged over Runs~(I--III) in comparison with the
               theoretical model~\cite{phili75} (solid line).}
\label{fig:energy2}
\end{figure}

In addition, the initial energy distributions of the protons and
deuterons can also be derived by analyzing the projections of the
two--dimensional distribution $(A_{jk})$ onto the $dE$ axis
$(A_{j})$ and the $E$ axis $(A_{k})$.
The equations for the $dE$ axis are
\begin{equation}
      \label{eq43}
      P(E_{p}^i / A_j ) = \frac{\tilde{S}(E_{p}^i )\sum\limits_{k = 1}^m
      {P^{MC}(A_{jk} / E_{p}^i )} }{\sum\limits_i {\tilde{S}(E_{p}^i ) \cdot
      \sum\limits_{k = 1}^m {P^{MC}(A_{jk} / E_{p}^i )} } }
\end{equation}
and 
\begin{equation}
      \label{eq42}
      N_{p}(\Delta E_{p},\Delta T) \cdot \tilde{S}(E_{p}^i ) =
      \frac{\sum\limits_{j = 1}^l {P(E_{p}^i / A_j ) \cdot (N_j)^{\exp }
      } }{P(A / E_{p}^i )} \, .
\end{equation}
Similar equations can be written for the $E$ axis.
Using the above equations, we obtain simulated values for the proton
and deuteron yields as measured by the Si($dE-E$) detectors.
Figure~\ref{fig:distr2} shows the projections of the experimental and
simulated ($dE-E$) distributions for protons and deuterons onto the
$E$ axis.

The mean proton $<S(E_p)>$ and deuteron $<S(E_d)>$ energy
distributions from Runs (I--III) are given in
Tables~\ref{tab:p-energy} and \ref{tab:d-energy}.
The mean values are also displayed in Fig.~\ref{fig:energy2}.
It is important to note that the distribution $S(E_{p})$ does
practically not depend on the form of the energy distribution
$S_o(E_{p})$ which is chosen for the first iteration.
Variation errors in the determination of $S(E_{p})$ fall within the
statistical errors of $(N_{jk})^{\exp}$.  Since Eqs.~(\ref{eq38}) and
(\ref{eq42}) (as well as the other projection) have an identical
solution, their comparison makes it possible to conclude, with an
accuracy determined by the statistics of the detected events, that
there are no systematic errors in the analysis of experimental data.

\begin{figure}[t]
      \includegraphics[angle=90,width=0.47\textwidth]{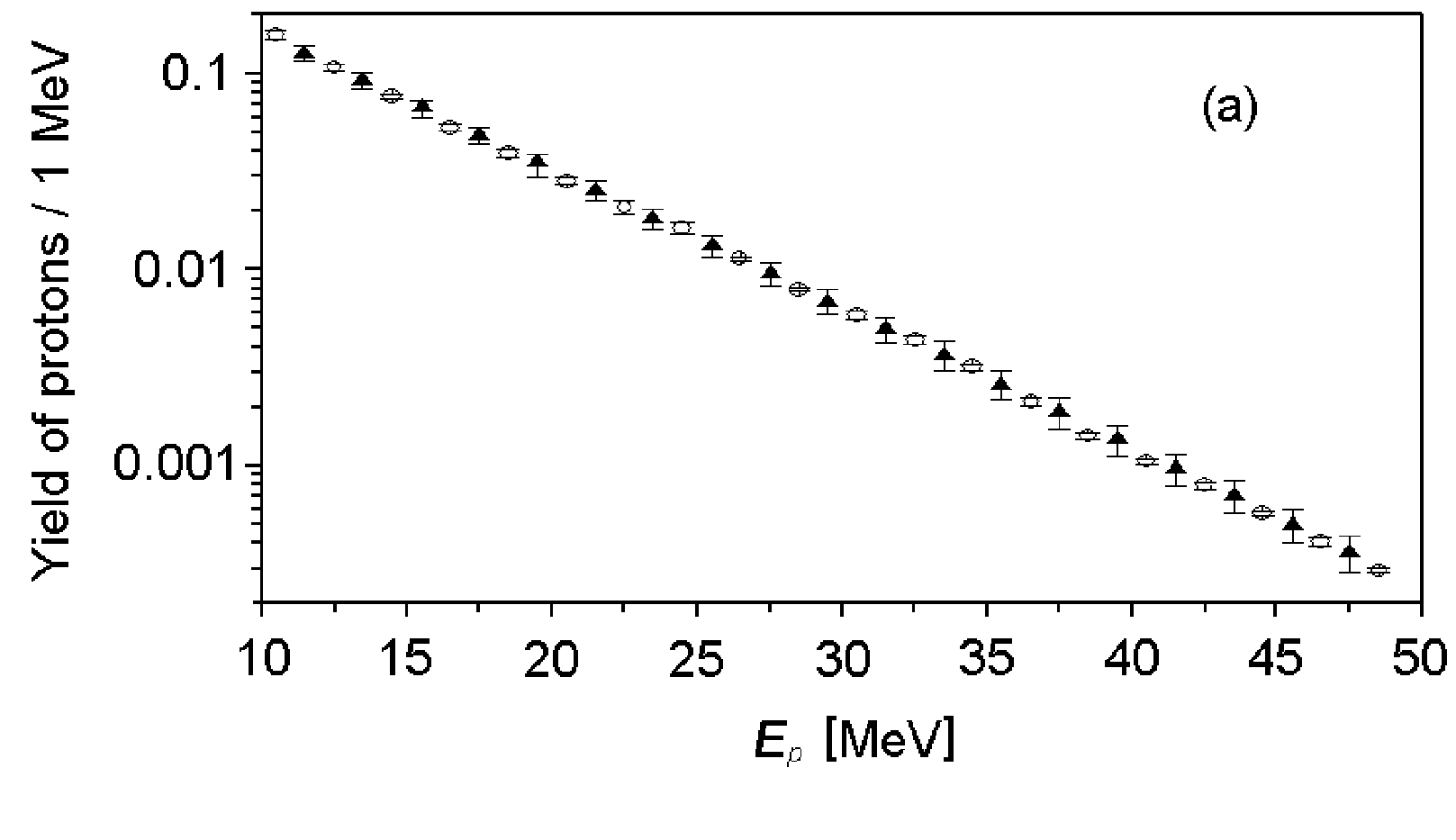}
      \includegraphics[angle=90,width=0.47\textwidth]{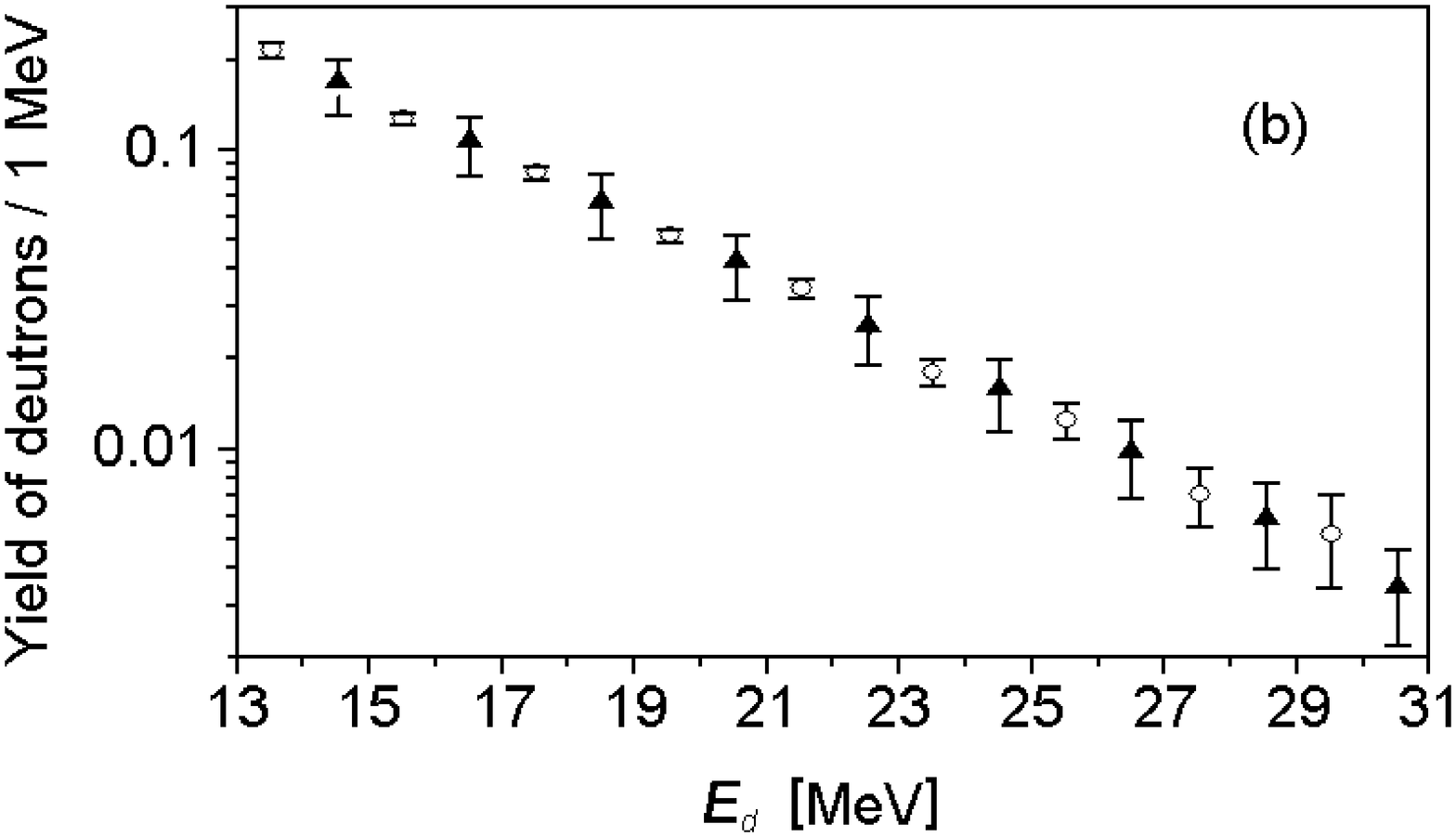}
      \caption{Comparison of the proton (a) and deuteron (b) energy
               distributions found by methods I (black triangles) and
               II (open circles) averaged over Runs~(I--III). For sake
               of visibility, we plotted both methods results
               alternatively. }
      \label{fig:compar}
\end{figure}

A comparison between the experimental energy distributions given in
Fig.~\ref{fig:energy2} and the energy distribution calculated by the
impulse approximation reveals some discrepancies of the same character
as in the method I analysis, as long as the interactions between the
reaction products~(\ref{eq1}) and~(\ref{eq2}) are considered and a
realistic Bonn B~\cite{skibi99} nucleon-nucleon potential is employed.

The capture rates $\lambda _\mathrm{cap}^{p} (\Delta E_{p} )$ as well
as the differential capture rates $d\lambda^p_\mathrm{cap} / dE_p $
which are found from Eqs.~(\ref{eq38}) and~(\ref{eq42}) using
Eqs.~(\ref{eq9}), (\ref{eq8}), and~(\ref{eq20}) are given in
Fig.~\ref{fig:capt1} for the protons and in Fig.~\ref{fig:capt2} for
the deuterons.

\section{Conclusions}
\label{sec:conclusions}

\begingroup
  \squeezetable
\begin{table}[t]
\begin{ruledtabular} 
      \caption{Relations $d\lambda _\mathrm{cap}^{p}(E_{p}) / dE_{p} $
               found by methods I and II and averaged over
               Runs~(I--III).  The proton energies $E_{p}$ correspond to
               the middle of the respective 1~MeV intervals.}
\label{tab:rela-p}
\begin{tabular}{ccc}
$E_p$ & \multicolumn{2}{c}{$<\frac{d\lambda _\mathrm{cap}^p }{dE_p }>$
    [MeV$^{-1}$s$^{-1}$] } \\ {}[MeV] & {Method I} & {Method II} \\
    &&\\ \hline 
10.5  & 5.49 (59)  & 5.77 (47)  \\
11.5  & 4.67 (51)  & 4.81 (35)  \\
12.5  & 3.98 (44)  & 3.95 (28)  \\
13.5  & 3.38 (38)  & 3.52 (25)  \\
14.5  & 2.88 (33)  & 2.81 (20)  \\
15.5  & 2.45 (29)  & 2.37 (17)  \\
16.5  & 2.08 (25)  & 1.93 (15)  \\
17.5  & 1.77 (22)  & 1.78 (13)  \\
18.5  & 1.51 (19)  & 1.44 (11)  \\
19.5  & 1.28 (16)  & 1.151 (95) \\
20.5  & 1.09 (14)  & 1.041 (88) \\
21.5  & 0.93 (12)  & 0.920 (77) \\
22.5  & 0.79 (11)  & 0.763 (71) \\
23.5  & 0.671 (92) & 0.675 (64) \\
24.5  & 0.570 (80) & 0.595 (56) \\
25.5  & 0.485 (69) & 0.549 (55) \\
26.5  & 0.412 (60) & 0.417 (28) \\
27.5  & 0.350 (52) & 0.343 (23) \\
28.5  & 0.298 (45) & 0.291 (19) \\
29.5  & 0.253 (39) & 0.249 (17) \\
30.5  & 0.215 (34) & 0.215 (16) \\
31.5  & 0.183 (29) & 0.194 (16) \\
32.5  & 0.155 (26) & 0.162 (13) \\
33.5  & 0.132 (22) & 0.139 (11) \\
34.5  & 0.112 (19) & 0.116 (9)  \\
35.5  & 0.095 (17) & 0.094 (7)  \\
36.5  & 0.081 (14) & 0.077 (6)  \\
37.5  & 0.069 (12) & 0.065 (5)  \\
38.5  & 0.058 (11) & 0.052 (4)  \\
39.5  & 0.050 (9)  & 0.044 (3)  \\
40.5  & 0.042 (8)  & 0.038 (3)  \\
41.5  & 0.036 (7)  & 0.034 (3)  \\
42.5  & 0.030 (6)  & 0.029~(2)  \\
43.5  & 0.026 (5)  & 0.024~(2)  \\
44.5  & 0.022 (4)  & 0.021~(2)  \\
45.5  & 0.019 (4)  & 0.018~(1)  \\
46.5  & 0.016 (3)  & 0.015~(1)  \\
47.5  & 0.013 (3)  & 0.013~(1)  \\
48.5  & 0.011~(2)  & 0.011~(1)  \\
\end{tabular}
\end{ruledtabular}
\end{table}
\endgroup

\begin{table}[t]
\begin{ruledtabular} 
      \caption{Relations $d\lambda _\mathrm{cap}^{d}(E_{d}) / dE_{d} $
               found by methods I and II and averaged over
               Runs~(I--III).  The deuteron energies $E_{d}$ correspond
               to the middle of the respective 1~MeV intervals.}
\label{tab:rela-d}
\begin{tabular}{ccc}
$E_d$ & \multicolumn{2}{c}{$<\frac{d\lambda _\mathrm{cap}^d }{dE_d }>$
[MeV$^{-1}$s$^{-1}$] } \\ {}[MeV] & {Method I} & {Method II} \\ &&\\
\hline 
13.5  & 4.46 (94)  & 4.74 (36)  \\
14.5  & 3.56 (77)  & 3.70 (26)  \\
15.5  & 2.84 (63)  & 2.81 (19)  \\
16.5  & 2.26 (51)  & 2.29 (16)  \\
17.5  & 1.80 (42)  & 1.84 (13)  \\
18.5  & 1.43 (34)  & 1.48 (11)  \\
19.5  & 1.13 (28)  & 1.141 (87) \\
20.5  & 0.90 (22)  & 0.933 (74) \\
21.5  & 0.70 (18)  & 0.756 (67) \\
22.5  & 0.55 (15)  & 0.550 (55) \\
23.5  & 0.43 (12)  & 0.397 (47) \\
24.5  & 0.340 (95) & 0.289 (44) \\
25.5  & 0.266 (76) & 0.272 (41) \\
26.5  & 0.207 (61) & 0.221 (37) \\
27.5  & 0.161 (49) & 0.156 (35) \\
28.5  & 0.124 (39) & 0.127 (42) \\
29.5  & 0.096 (31) & 0.114 (41) \\
30.5  & 0.074 (25) & 0.095 (31) \\
\end{tabular}
\end{ruledtabular}
\end{table}

The proton and deuteron energy distributions found by methods~I and~II
largely coincide within the measurement errors, which points to the
compatibility of the different approaches and to the absence of any
systematic errors which may have been neglected in the analysis of the
experimental data (see Fig.~\ref{fig:compar}).
However, the errors on $S(E_p)$ and $S(E_d)$ found by both methods are
different.
The analysis using method II gives a more
precise information about the proton and deuteron energy distributions than
method \@I.  
In method I, we compare using the numbers of detected events from a
($jk$) cell with similar MC simulated data.
Such numbers are the sums of the contributions from all $i$-th proton
energy subintervals $\Delta E^i_p$.
In method II, we have much deeper relations because the comparisons
are performed via Eq.~(\ref{eq38}) for each $i$--th subinterval
separately and all comparisons should be simultaneously satisfactory.

\begin{figure}[t]
      \includegraphics[angle=90,width=0.48\textwidth]{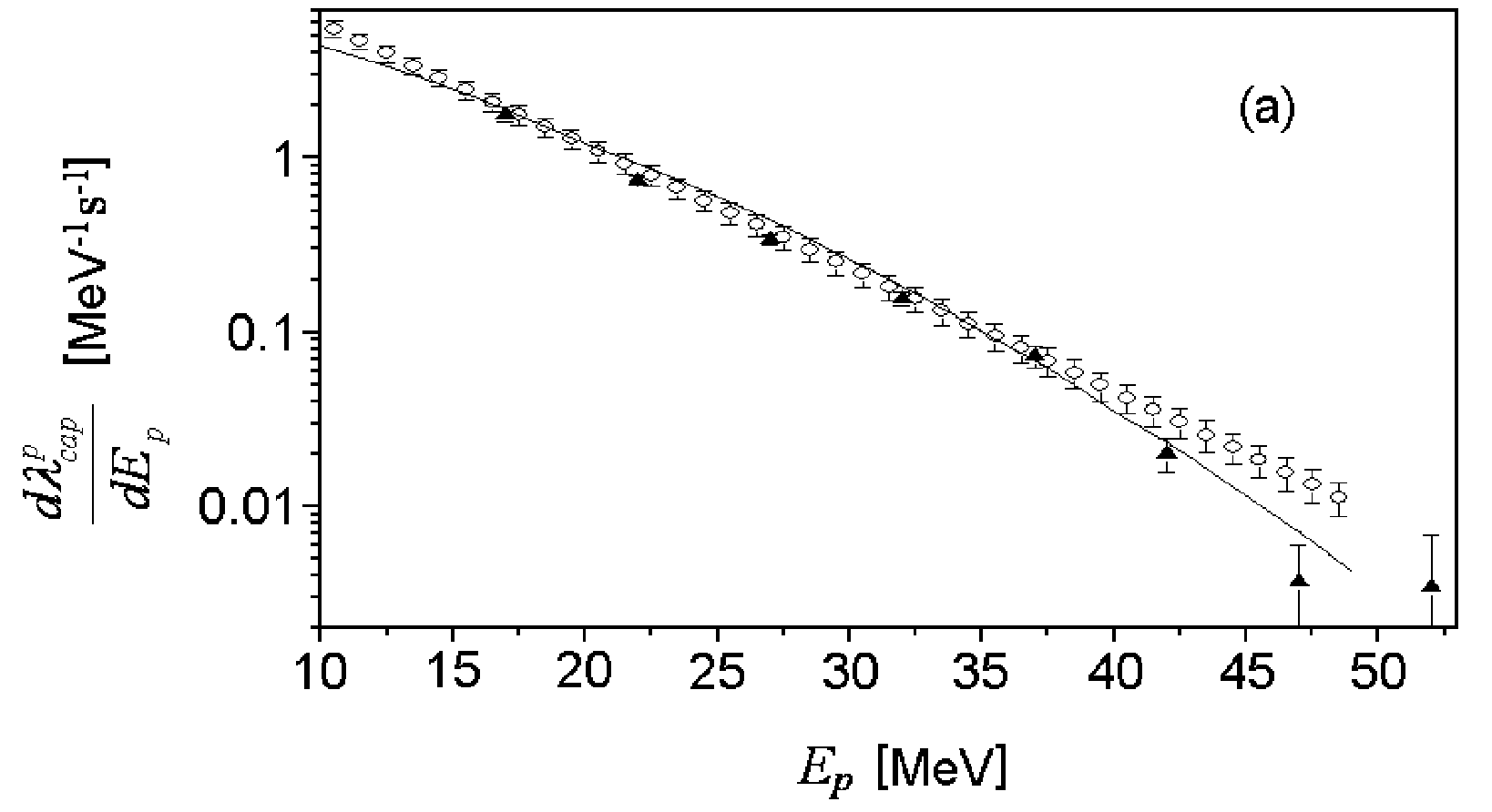}
      \includegraphics[angle=90,width=0.48\textwidth]{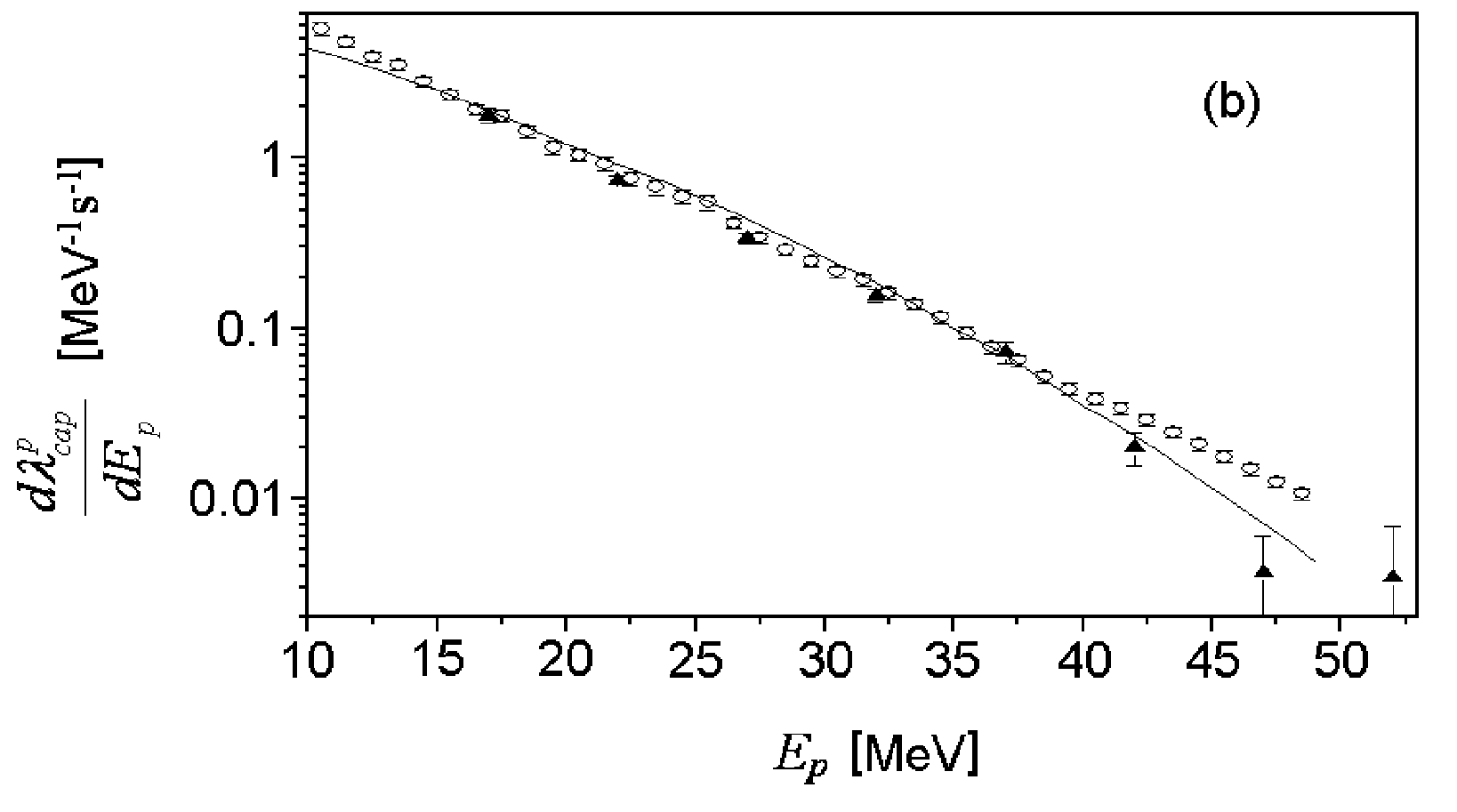}
      \caption{Differential rates $ d\lambda _\mathrm{cap}^p (E_p ) /
               dE_p $ (open circles) found by methods I (a) and II (b)
               averaged over Runs~(I--III).  Black triangles are the
               results of Refs.~\cite{kuhnx94,cummi92}; the solid line
               corresponds to the model~\cite{phili75}.  }
      \label{fig:capt1}
\end{figure}

Similar remarks hold for the differential capture rates $ d\lambda
_\mathrm{cap}^{p} (E_{p} ) / dE_{p} $ and $\lambda _\mathrm{cap}^{p}
(\Delta E_{p} )$, as seen in Fig.~\ref{fig:capt1}.
Tables~\ref{tab:rela-p} and~\ref{tab:rela-d} list the values of $<
d\lambda _\mathrm{cap}^{p} (E_{p} ) / dE_{p} > $ and $< d\lambda
_\mathrm{cap}^{d} (E_{d} ) / dE_{d} > $ found from the analysis of the
Runs~(I--III) with pure ${}^{3}\mathrm{He}$ and $\mathrm{D}_2 +
{}^{3}\mathrm{He}$ mixtures data by methods I and II\@.

\begin{table}[b]
\begin{ruledtabular} 
      \caption{Muon capture rates by ${}^{3}\mathrm{He}$ nucleus (in
               s$^{ - 1}$) followed by the proton and deuteron
               production following methods I and II\@.}
\label{tab:results}
\begin{tabular}{ccc}
Method & I & II \\ 
%       & [s$^{ - 1}$] & [s$^{ - 1}$] \\ 
\hline
$\lambda _\mathrm{cap}^p (10 \le E_{p} \le 49 \, \mbox{MeV})$
 & $36.7 \pm 1.2$ 
 & $36.8 \pm 0.8$  \\
$ \lambda _\mathrm{cap}^d (13 \le E_{p} \le 31 \, \mbox{MeV})$
 & $21.3 \pm 1.6$
 & $21.9 \pm 0.6$ \\
\end{tabular}
\end{ruledtabular}
\end{table}

The addition of the differential rates in Tables~\ref{tab:rela-p}
and~\ref{tab:rela-d} yields the muon capture rates by the
${}^{3}\mathrm{He}$ nucleus followed by proton and deuteron
production in the final state in the energy intervals $10 \le
\mathrm{E}_{p} \le 49$~MeV and $13 \le E_{d} \le 31$~MeV, respectively (see
Table~\ref{tab:results}).

\begin{figure}[t]
      \includegraphics[angle=90,width=0.48\textwidth]{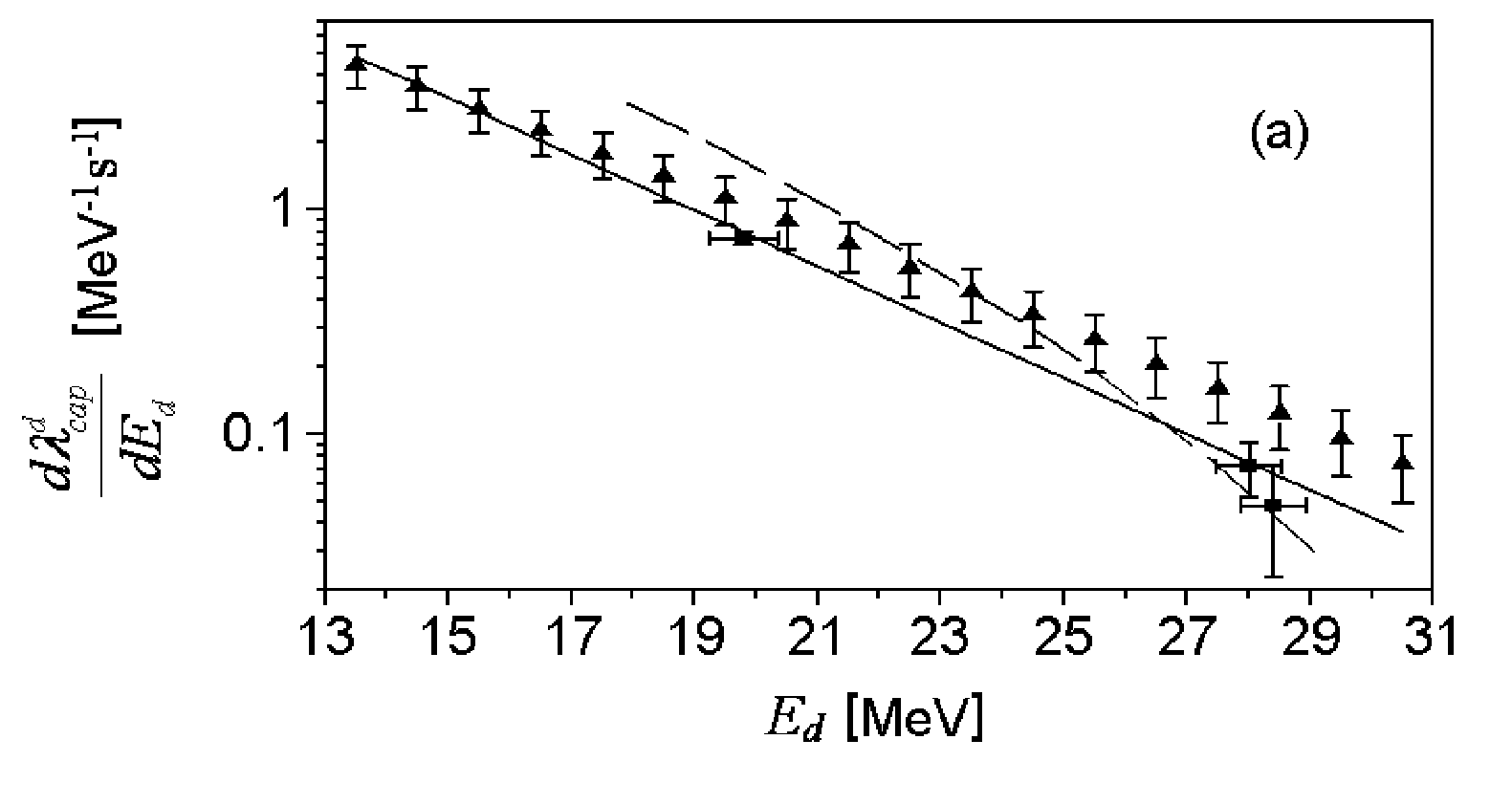}
      \includegraphics[angle=90,width=0.48\textwidth]{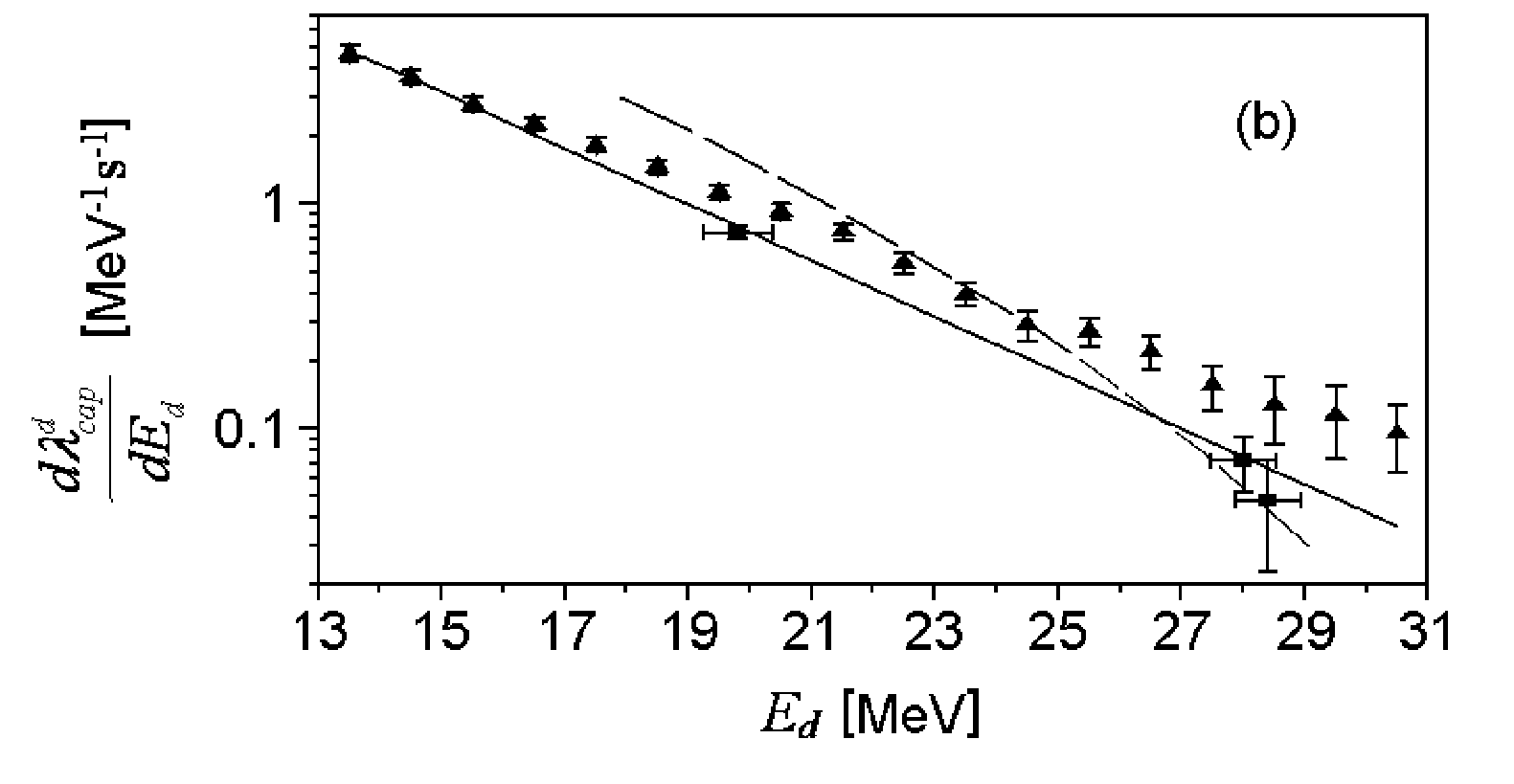}
      \caption{Differential rates $d\lambda _\mathrm{cap}^d (E_d ) /
               dE_d $ (black triangles) found by methods I (a) and II
               (b) and averaged over Runs~(I--III).  Black boxes are
               the results of Refs.~\cite{kuhnx94,cummi92}; the solid line
               corresponds to the model~\cite{phili75}; the dotted
               line is based on calculations from
               Ref.~\cite{skibi99}.}
      \label{fig:capt2}
\end{figure}

Looking more closely at Figs.~\ref{fig:capt1} and~\ref{fig:capt2}, our
results and their comparison with the experimental
data~\cite{kuhnx94,cummi92} and the
calculations~\cite{phili75,skibi99,glock96} indicate the following
results for the protons and deuterons.
Experimental (obtained by methods I and II) and calculated
differential rates $ d\lambda _\mathrm{cap}^p (E_p ) / dE_p $ and
$\lambda _\mathrm{cap}^{p} (\Delta E_{p} )$ for muon capture by the
${}^{3}\mathrm{He}$ nucleus followed by proton production in the
energy range $10 \le E_{p} \le 40$~MeV show quite good agreement both
in form and in magnitude.
The calculations were carried out in the simple plane--wave impulse
approximation (PWIA) with allowance made for final--state interaction
of reaction~(\ref{eq1}) and~(\ref{eq2}) products. 
However, there is a difference between the results of the present
paper and the calculations~\cite{phili75} for  proton energies $E_{p}
> 40$~MeV\@.

The measured dependence $ d\lambda _\mathrm{cap}^d (E_d ) / dE_d $
found by using methods I and II is quite well described by the
theoretical PWIA dependence~\cite{phili75} in the deuteron energy
ranges $13 \le E_{d} \le 20$~MeV (Method I) and $13 \le E_{d} \le
17$~MeV (Method II), respectively.
For deuteron energies  $E_{d} > 20$~MeV there is a noticeable
discrepancy between experiment and theory~\cite{phili75}.
The measured values of $ d\lambda _\mathrm{cap}^d (E_d )/ dE_d $ and
the PWIA calculations~\cite{skibi99} with the refined realistic NN
interaction potential (Bonn B) appreciably disagree over the entire
deuteron energy range.

Next, we can estimate the total capture rate (full energy range
$[0;\infty)$) using a simple extrapolation of our data at low energies
and a one--exponential weighted fit of the differential capture rate
in the full energy range.
Using the function 
\begin{equation}
      \label{eq200}
      \frac{d\lambda _\mathrm{cap}^p (E_p )}{dE_p} = H \cdot e^{- G
      \cdot E_p} \, ,
\end{equation}
where $H$ and $G$ are free parameters, we obtain the total capture
rate for the proton as their ratio
\begin{equation}
      \label{eq201}
      \lambda_\mathrm{cap}^p = \frac{H}{G} \, .
\end{equation}
Results for protons and deuterons, using both methods, are given in
Table~\ref{tab:rate}.
The summed rate $\lambda_\mathrm{cap}^p + \lambda_\mathrm{cap}^d$
(which corresponds to Eq.~(\ref{eq4a}) without the triton contribution) is also
compared to other experimental~\cite{zaimi63b,auerb65,maevx96} and
theoretical~\cite{yanox64,phili75,congl94} values.
Agreement between our results and previous ones is excellent.

\begin{table}[b]
\begin{ruledtabular}
      \caption{Total muon capture rate for reactions~(\ref{eq1})
               and~(\ref{eq2}). The results of this work is an
               estimation from both methods I (least--square) and II
               (Bayes theorem).}
\label{tab:rate}
\begin{tabular}{cccc}
Method & $\lambda_\mathrm{cap}^p$ & $\lambda_\mathrm{cap}^d$ &
$\lambda_\mathrm{cap}^p + \lambda_\mathrm{cap}^d$ \\
              & [s$^{-1}$]   & [s$^{-1}$] & [s$^{-1}$] \\ \hline
This work     &              &            &            \\
Least--square & $187\pm11$   & $491\pm125$ & $678\pm 126$ \\
Bayes         & $190\pm7$    & $497\pm 57$ & $687\pm60$ \\ \hline
Zaimidoroga~\cite{zaimi63b} &&            & $660 \pm 160$ \\
Auerbach~\cite{auerb65} &    &    & $665 \:\:^{+ \: 170}_{- \: 430}$\\
Maev~\cite{maevx96} &        &    & $720 \pm 70 $\\
Yano~\cite{yanox64} &        &            & 670 \\ 
Philips~\cite{phili75} & 209 & 414        & 623 \\
Congleton~\cite{congl94} &   &            & 650 \\
\end{tabular}
\end{ruledtabular}
\end{table}

An experimental determination of muon capture on ${}^{3}\mathrm{He}$
nuclei makes a study of electromagnetic and weak interactions of
elementary particles with 3N systems possible without introducing
uncertainties due to inadequate approximations of 3N states in the
analysis.
According to the theory, meson exchange currents must also be taken
into account in future analysis of experimental data.
As compared with Refs.~\cite{kuhnx94,cummi92}, this experiment
yields for the first time information on the ``softer'' region of
proton and deuteron energy spectra, which is more sensitive to the
theoretical models describing the final--state nucleon--nucleon
interactions.

Finally it should be mentioned that by increasing the efficiencies
of the proton and deuteron detection systems and their functional
capabilities, by decreasing the lower and increasing the upper
thresholds in the Si($dE-E$) telescopes, the above method will provide
precise information on the characteristics of muon capture by bound
few--nucleon systems.
It then becomes possible to verify various theoretical models of muon
capture by helium nuclei and to clarify the nature of discrepancies
between the results of the present paper and the experimental
data~\cite{kuhnx94,cummi92}.

\begin{acknowledgments}

The authors are grateful to Dr.~C.~Petitjean (PSI, Switzerland) and
C.~Donche--Gay for their support to this experiment, and to
Prof.~H.~Witala, Drs.~J.~Golak, and R.~Skibinski (Institute of
Physics, Jagiellonian University, Cracow, Poland) for helpful
discussions.
This work was supported by the Russian Foundation for Basic Research
(grant \#01--02--16483), the Swiss National Science Foundation and the
Paul Scherrer Institute.
\end{acknowledgments}

%\bibliography{mucf}

\end{document}